\documentclass[a4paper,fleqn,usenatbib]{mnras}

\usepackage{newtxtext,newtxmath}

\usepackage[T1]{fontenc}
\usepackage{ae,aecompl}

\usepackage{graphicx}	
\usepackage{amsmath}	
\usepackage{amssymb}	
\usepackage{bm}

\usepackage[dvipsnames]{xcolor}

\newcommand{\B}{\bm{B}}
\newcommand{\V}{\bm{v}}
\newcommand{\U}{\bm{u}}
\newcommand{\J}{\bm{J}}

\newcommand{\rp}{r_{\rm p}}
\newcommand{\wc}{w_{\rm c}}
\newcommand{\au}{{\, \rm au}}

\title[Low mass planet dead zone migration]{Low mass planet migration in magnetically torqued dead zones -- I. Static migration  torque}

\author[C.~P.~McNally et al.]{
Colin P.~McNally,$^{1,2}$\thanks{E-mail: c.mcnally@qmul.ac.uk (CPM)}
Richard P.~Nelson,$^{1,2}$
Sijme-Jan Paardekooper,$^{1,3}$\newauthor
Oliver Gressel,$^{4,2}$
and Wladimir Lyra$^{5,6,2}$ \\
$^{1}$Astronomy Unit, School of Physics and Astronomy, Queen Mary University of London, London E1 4NS, UK\\
$^{2}$Kavli Institute for Theoretical Physics, University of California Santa Barbara, CA 93106, USA\\
$^{3}$DAMTP, University of Cambridge, Wilberforce Road, Cambridge CB3 0WA, UK\\
$^{4}$Niels Bohr International Academy, The Niels Bohr Institute, Blegdamsvej 17, DK-2100, Copenhagen \O, Denmark\\
$^{5}$Department of Physics and Astronomy, California State University Northridge, 18111 Nordhoff St, Northridge, CA 91330, USA\\
$^{6}$Jet Propulsion Laboratory, California Institute of Technology, 4800 Oak Grove Drive, Pasadena, CA 91109, USA
}

\date{Accepted 2017 August 15. Received 2017 August 15; in original form 2017 July 2}

\pubyear{2017}

\begin{document}
\label{firstpage}
\pagerange{\pageref{firstpage}--\pageref{lastpage}}
\maketitle

\begin{abstract}
Motivated by models suggesting that the inner planet forming regions of protoplanetary discs
are predominantly lacking in viscosity-inducing turbulence, and are possibly threaded by 
Hall-effect generated large-scale horizontal magnetic fields,
we examine the dynamics of the corotation region of a low-mass planet in such an environment.
The corotation torque in an inviscid, isothermal, dead zone ought to saturate, with the libration 
region becoming both symmetrical and of a uniform vortensity, 
leading to fast inward migration driven by the Lindblad torques alone. 
However, in such a low viscosity situation, 
the material on librating streamlines essentially preserves its vortensity. 
If there is relative radial motion between the disc gas and the planet, the librating streamlines will no longer be symmetrical. 
Hence, if the gas is torqued by a large scale magnetic field so that it
undergoes a net inflow or outflow past the planet, driving evolution of the vortensity and inducing 
asymmetry of the corotation region, the corotation torque can grow, leading to a positive torque.
In this paper we treat this effect by applying a symmetry argument 
to the previously studied case of a migrating planet in an inviscid disc.
Our results show that the corotation torque due to a laminar Hall-induced magnetic field in a 
dead zone behaves quite differently from that studied previously for a viscous disc. 
Furthermore, the magnetic field induced corotation torque and the dynamical 
corotation torque in a low viscosity disc can be regarded as one unified effect.
\end{abstract}




\section{Introduction}
The gravitational interaction between a forming planet and its nascent protoplanetary disc becomes important as the planet approaches the Earth's mass, leading to angular momentum exchange and migration of the planet. For a non-gap forming low mass planet, the net migration torque has contributions from Lindblad and corotation torques that normally have opposite signs \citep[see][for recent reviews]{2012ARA&A..50..211K,2014prpl.conf..667B}. The Lindblad contribution arises from the excitation of spiral density waves, and almost always drives inwards migration \citep[e.g.][]{1980ApJ...241..425G}. The corotation torque arises from coorbital material undergoing horseshoe orbits, and generally drives outwards migration in the presence of negative radial entropy and/or vortensity gradients in the disc \citep{1991LPI....22.1463W, 2001ApJ...558..453M, 2008A&A...478..245P, 2008ApJ...672.1054B}. To date, migration theory has largely considered viscous discs, and in the absence of sufficient disc viscosity it is known that the corotation torque saturates (i.e. switches off), leaving only the Lindblad torque to drive rapid inwards migration \citep{2011MNRAS.410..293P}.
 
When taking into account non-ideal magnetohydrodynamics, recent studies have shown that protoplanetary discs do not behave in detail like viscous $\alpha$-discs.
Models suggest that if the disc is threaded by a weak net vertical field, ambipolar and Ohmic diffusion yield a largely magnetically dead disc, 
with an essentially laminar flow, and accretion primarily driven by a disc wind launched from a narrow, highly ionized region near the disc surfaces \citep{2013ApJ...769...76B,2015ApJ...801...84G}.
In addition, in the dead zone located at stellocentric distances $0.5 \lesssim R \lesssim 10 \au$, the Hall effect can be large and active when the background vertical field is aligned with the rotation of the disc \citep{1999MNRAS.303..239W,2008MNRAS.385.2269P}, 
creating significant laminar horizontal magnetic fields that generate an associated radial Maxwell stress \citep{2013ApJ...769...76B,2014A&A...566A..56L,2017A&A...600A..75B}. In this work, we consider the problem of low mass planet migration in a wind-driven protoplanetary disc 
\citep{
2013ApJ...769...76B,
2013ApJ...772...96B,
2014ApJ...791...73B,
2014ApJ...791..137B,
2014A&A...566A..56L,
2015ApJ...798...84B,
2015MNRAS.454.1117S,
2016ApJ...818..152B,
2016ApJ...821...80B,
2017A&A...600A..75B}.
Given that the wind is launched from high altitudes in the disc, it is unlikely that this will have a significant effect on the migration of a low mass planet located at the midplane, so we are specifically interested in the scenario when the Hall effect, by enabling the Hall-shear instability \citep{2008MNRAS.385.1494K,2013MNRAS.434.2295K} in the main planet forming region of a protoplanetary disc, leads to the generation of a midplane laminar Hall stress \citep{2014ApJ...791..137B,2014A&A...566A..56L,2016ApJ...819...68X,2017A&A...600A..75B}. 
While the magneto-thermal wind is driving accretion at the ionized surface,
the Maxwell stress due to the laminar spiral fields generated through the Hall-shear instability can drive radial flows at the midplane, and hence affect the corotation torque acting on a low mass planet.

Ignoring for a moment the effect of Hall-generated fields, when the dead zone is very magnetically decoupled (low magnetic Reynolds and Elsasser numbers), 
and where it  has a long enough cooling time to be 
stable against the Goldreich-Schubert-Fricke instability (or vertical shear instability, VSI) \citep{1967ApJ...150..571G, 1968ZA.....68..317F, 1998MNRAS.294..399U, 2013MNRAS.435.2610N} 
{ and lacks a sufficiently favourable entropy gradient to drive the convective overstability \citep{2014ApJ...788...21K,2014ApJ...789...77L} }
the dead zone has very low effective viscosity.
This combination of conditions would be expected in the inner part of the disc, at $\sim 0.5-10 \ \mathrm{au}$ \citep{2015MNRAS.454.1117S,2015ApJ...811...17L}, 
although variations between discs are likely to occur due to varying ionization chemistry and thermodynamics.
If a low mass planet is introduced to this part of a disc the properties of the material trapped on librating horseshoe streamlines in the corotation region have two important attributes.
First, the material becomes well mixed, such that any initial radial vortensity gradient is removed on a few libration times \citep{1991LPI....22.1463W,1992LPI....23.1491W}, and secondly,
the libration region has a long memory of the initial vortensity contained within it as no viscosity mixes the vortensity with the surrounding disc.
For a planet in a fixed position relative to a radially unmoving disc, the librating streamlines are symmetric in azimuth about the planet's position,
and the orbital phase-mixing of the librating material results in a saturation, or cancelling to zero, of the torque due to the corotation region.
However, if the corotation region is not symmetrical about the planet position, the asymmetry of the material on the leading and trailing horseshoe turns
results in a corotation torque, even when the librating material is well phase mixed.
Such a situation can be realized with a migrating planet, where the motion of the planet and history 
of the corotation region result in a dynamical corotation torque \citep{2014MNRAS.444.2031P}, even in an inviscid disc.
\footnote{This differs from the  hypothesis suggested by 
\citep{1991LPI....22.1463W,
1992LPI....23.1491W} where the vortensity gradient was conjectured to be maintained
if the planet were to drift across the corotation region in a time-scale short compared to the libration time.}

In this paper, we consider the effect of a laminar Hall-effect generated spiral magnetic field 
in the dead zone, and its associated Maxwell stress, on the corotation torque acting on a non-migrating planet in an inviscid disc. We expect the Maxwell stress to generate a radial flow in the disc, which may act similarly to the above mentioned moving planet in an inviscid disc, by generating a sustained corotation torque. Prior to undertaking a full analysis of the problem in three-dimensions with all the non-ideal magnetohydrodynamic (MHD) effects included, our approach here is to explore this scenario using a reduced two-dimensional model in which a steady spiral magnetic field is obtained in the disc midplane by a balance between winding of radial magnetic field and field diffusion due to Ohmic resistivity.

The paper is organized as follows. In Section~\ref{sec:analytical} we derive an analytical approximation for the corotation torque,
by first constructing a reduced model of the Hall-stress dominated dead zone, then after considering the 
relative magnitudes of the relevant length and time-scales we derive a formula for the evolution of the corotation torque.
In Section~\ref{sec:numericalmodels} we describe numerical models and experiments that verify 
the assumptions made in the analytical treatment and the torque formula derived. We discuss our results in Section~\ref{sec:discussion} and draw conclusions in Section~\ref{sec:conclusions}.

\section{Analytical Treatment} 
\label{sec:analytical}

Our model for the corotation torque requires two parts. 
First, an appropriate reduced model of a protoplanetary disc dead zone where the Hall-shear instability is maintaining a laminar spiral magnetic field that drives radial flow of the gas. Secondly, an evaluation of the evolution of the corotation torque given this background disc model.
The parameters of the reduced disc model inform the construction of the corotation torque approximation.

\subsection{Reduced model of dead zone}
\label{sec:discmodel}
Our concept of a Hall-effect modified dead zone draws on local \citep{2014ApJ...791..137B,2014A&A...566A..56L,2016ApJ...819...68X} 
and global models \citep{2017A&A...600A..75B} of
wind-driven protoplanetary discs including the Hall effect.
Where the disc rotation vector $\bm{\Omega}$ and the background magnetic field $\bm{B}_0$ are aligned, or
$\bm{\Omega} \cdot \bm{B}_0 >0$, a strong laminar field can develop, and as a prototypical example we take 
that realized in \citet{2016ApJ...819...68X} their fig.~5 or similarly \citet{2014A&A...566A..56L} their fig.~9.
Importantly, the field is so dominated by the azimuthal and radial components that it lies nearly purely in the plane of the disc.
Moreover, the radial component of the field is nearly constant with height in the central two scale heights of the disc.
Thus, to reduce the model to a two-dimensional radial-azimuthal one, we take that the radial field is simply imposed 
by the Hall-shear instability acting where the magnetic field has important variations above the midplane.
Also, being in two dimensions and requiring that $\nabla\cdot B=0$  dictates that the radial magnetic field obey  $B_r \propto r^{-1}$.
The Hall stress need not be the main driver of accretion, or even driving in the same direction as the net accretion, since the wind contribution at the disc surface may be dominant.
As the main action of the Hall effect in this weakly-coupled regime is in generating the radial field, and in this instability process the unstable wave vectors are directed vertically, we proceed to drop the Hall terms, and retain only Ohmic diffusion, 
which limits how the imposed radial field can wind up to form a spiral.

Thus, having chosen a two dimensional radial-azimuthal configuration, assuming a thin disc, and imposing a radial magnetic field, 
we can proceed to construct an approximate power-law solution for an equilibrium steady state.
In doing this, we will impose the constraint that the accretion rate is constant with radius.
In the thin disc, the non-dimensional scale height $h\equiv H/r$ is small  $h \ll 1$, and so the azimuthal velocity of the 
gas is approximately Keplerian.
We thus assume Keplerian rotation  { with the azimuthal velocity} $\V_K = \Omega_0(r/r_0)^{-1/2}$ { where $\Omega_0$ is the Keplerian angular velocity at radius $r_0$}, and a power law in surface density {$\Sigma$ given by }
\begin{align}
\Sigma = \Sigma_0 \left(\frac{r}{r_0}\right)^{-\alpha}\, ,
\end{align}
{ where $r$ is the cylindrical radius and $r_0$ is a reference radius.}
We can then take the induction equation as
\begin{align}
 \frac{\partial \B}{\partial t} &= \nabla \times\left(\V \times \B -  \eta \J\right)\\
&\J \equiv \frac{1}{\mu_0} \nabla \times \B
\end{align}
where { $\B$ is the magnetic field, $t$ is time, $\V$ is the gas velocity, $\J$ is the current density,} $\eta$ is the Ohmic resistivity and $\mu_0$ is the vacuum permeability (note this is the SI formulation, Ohmic diffusivity is $\eta/\mu_0$),
and solve for a steady state $ {\partial \B}/{\partial t} =0$, with an imposed Keplerian flow.
Requiring a constant accretion rate with radius gives
\begin{align}
B_r &= B_0 \left(\frac{r}{r_0}\right)^{-1} \label{eq:bric}\\
B_\phi &= -2 B_0 \Omega_0 r_0^2 \frac{\mu_0}{\eta} \left(\frac{r}{r_0}\right)^{-1/2} \label{eq:bphiic}
\end{align}
with the radial velocity as
\begin{align}
v_r &= -\frac{2 B_0^2 r_0^2}{ \eta \rho L_z r}\, ,\label{eq:vr}
\end{align}
where $\Sigma =  \rho L_z$, $\rho$ is the midplane density, and we take $L_z$ to be a constant in these models.
In practice, this approximation is neither exact or a numerical equilibrium, 
and we allow the disc to relax from this initial condition 
to a numerical equilibrium before introducing a planet.
Further details of such procedures are given in Section~\ref{sec:numericalmethods}.

We comment here that most of our production run models that are presented in Section~\ref{sec:numericalmodels} will be run with a body force that replaces and mimics the Lorentz force due to the magnetic field, and do not use this magnetic field configuration explicitly.
The explicit use of this approximate magnetic field solution in this work is to establish 
that although it is physically self-consistent, the feedback of the planet-perturbed 
flow on to the magnetic field configuration is so weak that the induction equation 
can consistently be dropped from our models, since there is essentially no time evolution of the magnetic field once the equilibrium described by equations~(\ref{eq:bric}) and (\ref{eq:bphiic}) has been established.

Although this magnetic field configuration is very specific,
 we will show theoretically in Section~\ref{sec:scales} and numerically in Section~\ref{sec:equiv} that the large resistivity of the 
Ohmic dead zone implies that the corotation torque only depends on the 
effective body force which changes the angular momentum of the disc gas, not the details of the 
magnetic field configuration itself.
Hence, even though our models are constructed following a spiral magnetic field, the results depend only on the net force exerted on the fluid.
Thus, if the radial flow were instead driven by vertical Maxwell stresses ($\langle B_z B_\phi \rangle$) but still 
resulted in laminar inflow or outflow throughout the disc column, 
the results in this work will still apply.

\subsection{Analytical theory of the corotation torque}

\begin{figure}
\includegraphics[width=\columnwidth]{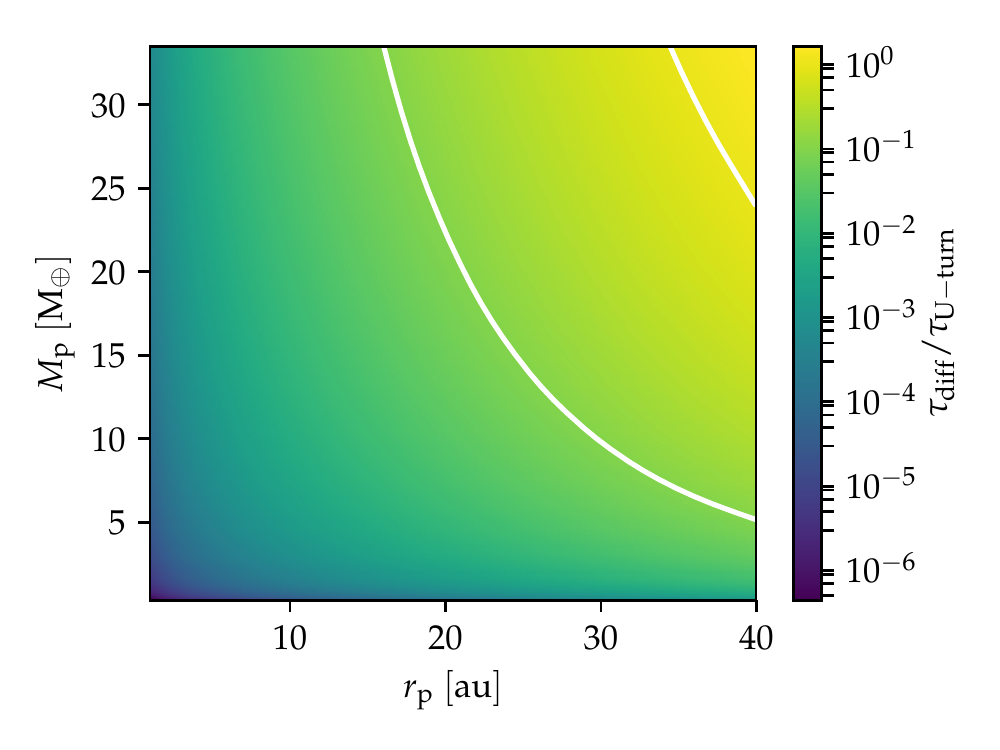}
\caption{Illustrative variation of $\tau_{\rm diff} / \tau_{\rm U-turn}$ (shown logarithmically) for the representative disc model described in the text. The two white contours show $\tau_{\rm diff} / \tau_{\rm U-turn}=0.1 ,\ 1$.}
\label{fig:tdiff-tuturn}
\end{figure}

\begin{figure}
\includegraphics[width=0.8\columnwidth]{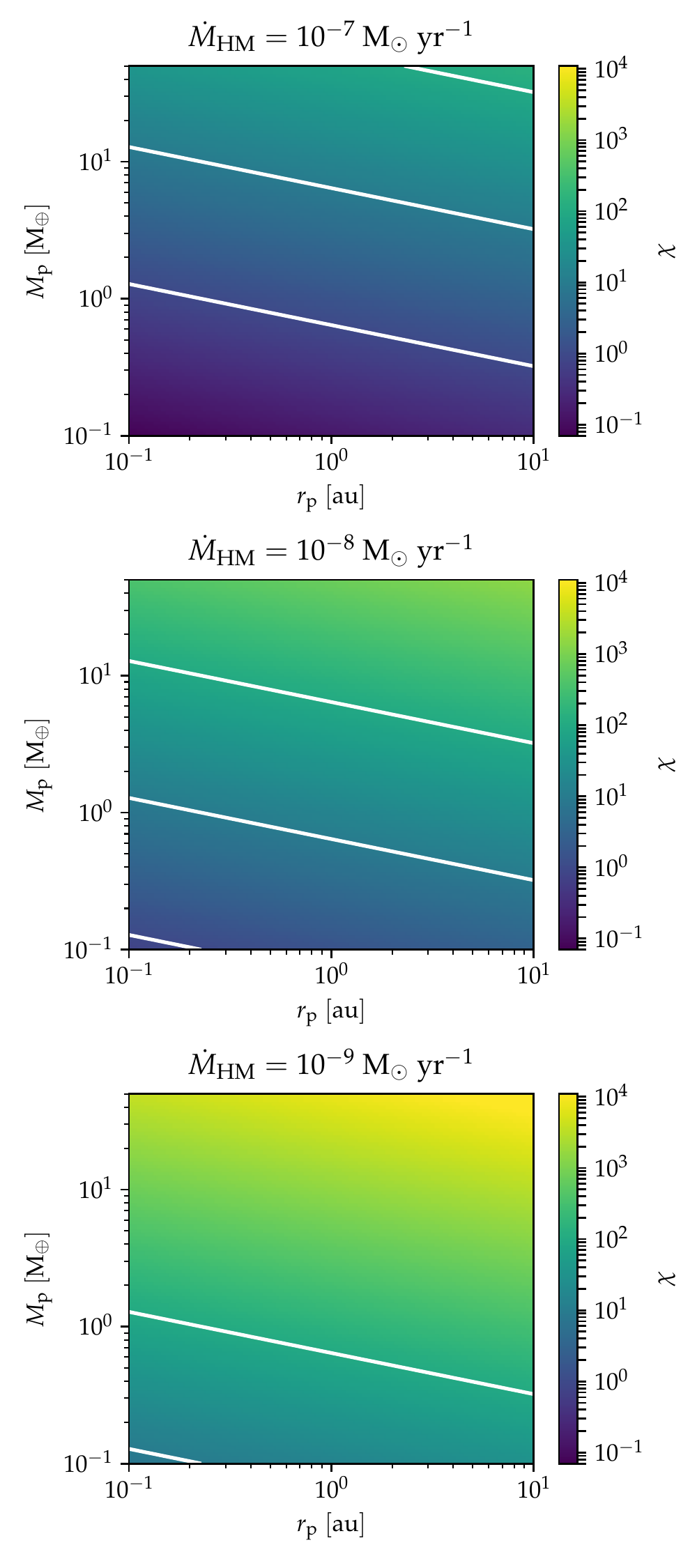}
\caption{Illustrative variation of the $\chi$ parameter for three discs with $\Sigma=1700 (r/(1\ {\rm au}))^{-1}\ {\rm g\, cm^{-2}}$, flaring (scale height variation) $h = 0.05 r_{\rm p}^{1.1}$, and varying 
mid-plane laminar Hall-stress driven midplane accretion rates $\dot{M}_{\rm HM}$. The three white contours show $\chi=1 ,\ 10,\ 100$.}
\label{fig:mdotchi}
\end{figure}

Given the description of the two dimensional equilibrium disc model established in the previous subsection, we can derive a theory for the corotation torque experienced by a low-mass planet. 
First the appropriate regime for the approximations will be established by examining the relevant 
length and time-scales, and then the detailed evolution of the corotation region and torque will be derived.

\subsubsection{Length and time-scales}
\label{sec:scales}

For a low mass planet, the half-width of the corotation region is reasonably approximated as
\begin{align}
x_s \approx 1.2 r_p \sqrt{q/h}
\end{align}
{ where $r_p$ is the planet's radial position and $q$ is the planet-star mass ratio, or equivalently the mass of the planet in units of the mass of the central star} \citep{2006ApJ...652..730M}.
We take as the characteristic time-scale of the libration region the time-scale for a fluid element on the separatrix
at distance $x_s$ away from the planet to make  one libration
\begin{align}
\tau_{\rm lib} & = \frac{4\pi}{x_s |d\Omega/dr|} = \frac{8\pi \rp}{3 \Omega_p x_s}\, ,
\end{align}
{ where $\Omega$ is the angular velocity of Keplerian orbits and $\Omega_p$ is the orbital velocity at the planet position.} 
The approximate diffusion time-scale of magnetic fields across the corotation region is
\begin{align}
\tau_{\rm diff} &= \frac{x_s^2}{\eta / \mu_0}.
\end{align}
The U-turn, or horseshoe-turn, time-scale is approximately \citep{2008ApJ...672.1054B}
\begin{align}
\tau_{\rm U-turn} &\approx h \tau_{\rm lib}.
\end{align}
As the inner dead zone has overwhelmingly large Ohmic diffusion \citep[e.g.][]{2015ApJ...801...84G}, we work in the limit $\tau_\text{diff} \ll \tau_\text{U-turn}$. 
Thus, we treat the magnetic field as a constant, with no feedback on its 
configuration from the flow around the planet.
This is in contrast to \citet{2013MNRAS.430.1764G} where the regime considered  for low mass migration in a weakly magnetized laminar disc was $\tau_\text{U-turn} <\tau_\text{diff}<\tau_\text{lib}/2$. 
Similarly, torques related to the magnetic resonances like those 
seen in laminar, magnetized, but well-coupled
discs \citep{2003MNRAS.341.1157T,2005MNRAS.363..943F} are not expected, as MHD waves are 
strongly damped by the large Ohmic diffusivity of the dead zone.

Comparing the two time-scales $\tau_\text{diff}$ and $\tau_\text{U-turn}$, one can find the critical value of resistivity for this ordering of time-scales to apply as being
\begin{align}
(\eta/\mu_0)_\text{crit} &\approx 3\times 10^{-5}  \frac{(q/2\times10^{-5})^{3/2}}{(h/0.05)^{5/2}} \Omega_0 r_0^2 \, .
\end{align}
Using the chemical network and ionization sources described in detail in \citet{2011MNRAS.415.3291G,2012MNRAS.422.1140G,2013ApJ...779...59G}, based originally on \texttt{model 4} in \cite{2006A&A...445..205I}, we have computed the equilibrium ionization fraction, and resulting resistivity profile, in a representative disc model, from which we have calculated the ratio $\tau_{\rm diff}/\tau_{\rm U-turn}$ as a function of planet mass and orbital location. The results are shown in Fig.~\ref{fig:tdiff-tuturn}, and clearly demonstrate that for low mass planets we expect to be comfortably in the regime $\tau_{\rm diff}/\tau_{\rm U-turn} < 1$. The disc model adopted is $\Sigma(R) = \Sigma_0 R^{-1/2}$, normalized such that $\Sigma(5\au)=150$ g cm$^{-2}$, and $H/R=0.05$ throughout the disc, with a dust-to-gas ratio of $10^{-3}$ in well-mixed small micron-sized grains.

In comparison, the critical resistivity level for MRI activity, defining the edge of the dead zone is given by the Ohmic Elsasser number
\begin{align}
\Lambda &\equiv \frac{ v_A^2 }{ (\eta/\mu_0) \Omega_0} \\
&= \frac{B/\sqrt{\mu_0 \rho} }{ (\eta/\mu_0) \Omega_0} 
\end{align}
where the strength of the magnetic field can be parametrized in terms of the plasma beta
\begin{align}
\beta &\equiv \frac{P}{B^2/(2\mu_0) } = \frac{2 c_s^2}{v_A^2}
\end{align}
which gives
\begin{align}
(\eta/\mu_0)_{\rm crit,MRI} &\approx \frac{c_s^2}{\beta \Omega_0} \approx 10^{-5} \left(\frac{ (c_s/0.05)^2 }{ (\beta/100)  } \right)\Omega_0 r_0^2.
\end{align}
Hence, we find that reasonable disc models with properties similar to the minimum mass solar nebula are very clearly in the regime $\Lambda < 1$ and $\tau_{\rm diff}/\tau_{\rm U-turn} < 1$ for low mass planets, as required by our model assumptions.

We define a time-scale, $\tau_{\rm f}$, which is approximately the time for a fluid element to 
move across the radial extent of the corotation region due to the radial flow at the midplane, as
\begin{align}
\tau_{\rm f} &\equiv \frac{2 x_s}{-v_r},
\end{align}
a sign convention chosen so that inward flow gives positive flushing times. We expect the controlling parameter for the modification of the horseshoe torque to be the ratio of time-scales between the libration time and
the time for the torque to flush the disc material past the planet.
We give this dimensionless parameter the symbol $\chi$, and define it by 
\begin{align}
\chi \equiv \frac{\tau_{\rm f}}{\tau_{\rm lib}} &= \frac{3 (1.2^2) }{8 \pi  } \frac{\Omega_0  \eta_0 \rho } { B_0^2 } \frac{q}{h}
\end{align}
and note it can be expressed in terms of the parameters of the planet and our disc model.
The definition of $\tau_{\rm f}$ is such that $\chi$ is positive for disc material flowing 
radially inwards with respect to the planet, and negative for disc material flowing
radially outwards with respect to the planet.

To describe an illustrative range of values of $\chi$ for different discs and planets,
in Fig.~\ref{fig:mdotchi} we show the values of $\chi$ which would occur in an example disc with
surface density $\Sigma=1700 (r/(1\ {\rm au}))^{-1}\ {\rm g\, cm^{-2}}$, 
flaring (scale height variation) $h = 0.05 r_{\rm p}^{1.1}$ and varying values of the midplane laminar stress driven mass accretion rate
which we denote $\dot{M}_{\rm HM}$.
The overall accretion rate on to the star in steady-state will be at least this value, 
and greater if a disc wind drives accretion at the surface of the disc.
Alternately, we can turn to local shearing-box models including Ohmic and Ambipolar diffusion and the Hall effect, such as those in
\citet{2016ApJ...819...68X} their Table~1 and assume an Shakura-Sunyaev $\alpha$ scaling for the stress with radius. 
For a planet with mass $q=5\times10^{-6}$ at $1\ {\rm au}$ such assumptions yield $\chi=1.5 (\alpha_{\rm HM}/10^{-3})$ where $\alpha_{\rm HM}$ 
is the radial Maxwell stress of the laminar magnetic field expressed as a Shakura-Sunyaev $\alpha$.
As this $\alpha_{\rm HM}$ is found to be on the order of $10^{-3}-10^{-4}$ in \citet{2016ApJ...819...68X}  it is reasonable to expect significant values for $\chi$ in such discs.

This time-scale, $\tau_{\rm f}$, has a similar nature to the time-scale relevant in dynamical corotation torques, that is the time for a 
planet to migrate across the radial length equal to the width of the corotation region.
This has figured prominently in the analysis of the corotation torque on migrating planets in
 \citet{2003ApJ...588..494M} and \citet{2014MNRAS.444.2031P}. Indeed, from this same approach we can now derive 
 an analytical prediction of the corotation torque when disc material is flowing past a non-migrating planet.

\subsubsection{Torque formula}
\label{sec:torqueformula}

\begin{figure}
\includegraphics[height=\columnwidth,angle=-90]{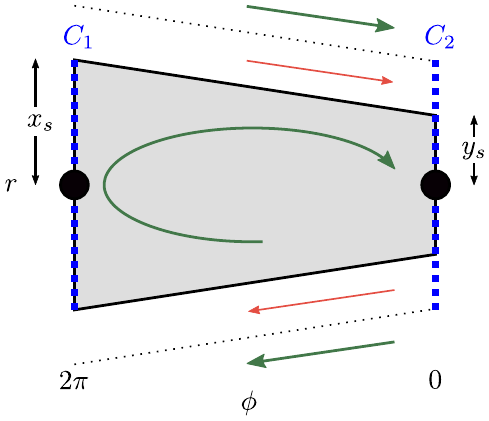}
\caption{Schematic diagram of the integration contours used in equation~(\ref{eq:genhs}), with the $\chi>0$ case shown in particular so that the asymmetry of the libration region (grey shading) is driven by the magnetic field induced gas radial inflow.
The flow is shown in the frame corotating with the planet. 
Shown in blue dashes are the cuts just in front and behind of the planet across the corotation region with half-width $x_s$, 
which behind  the planet ($C_1$) includes material in the librating streamlines (grey shading), 
and in front  of the planet ($C_2$) cuts through material inside the librating streamlines 
and disc gas on flow-through streamlines which only have a single close encounter with the planet (thin red arrow, inside dotted black line).}
\label{fig:integral_contours}
\end{figure}

We restrict our attention to the case $\chi>1$, which is analogous to slow migration for a moving planet in an untorqued disc.
If instead  $\chi\ll 1$ the corotation torque modification would be like the one envisioned by \citet{1991LPI....22.1463W,
1992LPI....23.1491W}.

The dragging of disc gas by the Lorentz force will have two effects. It will deform the librating streamlines in the corotation region, and it will allow some disc gas to flow past the planet from the outer disc to the inner disc, experiencing a single encounter with the planet as it does so.
In the $\chi \gg 1$ regime (equivalent to slow migration), 
we can approximately describe the asymmetry of the librating streamlines by how far inwards we expect the material initially passing the planet at $x_s$ to move inwards under the influence of the Lorentz force by the time it next encounters the planet (after $\tau_{\rm lib}/2$):
\begin{align}
y_s = x_s + \frac{\tau_{\rm lib}}{2} v_r
\end{align}
where $x_s$ is the earlier defined half-width of the corotation region, and thus $y_s$ is the width of the narrower end of the librating streamlines.
The narrow end of the island of librating streamlines is located azimuthally in front of the planet in its orbit for $\chi>0$, but is behind it for $\chi<0$ as the Lorentz force is dragging the disc gas outwards in that case.
Then, following
\citet{2003ApJ...588..494M,2009MNRAS.394.2283P,2014MNRAS.444.2031P}, we can represent the corotation torque quite 
generally as a horseshoe torque.
A horseshoe torque derives from considering the asymmetry of the horseshoe turns at each end of the librating streamlines, 
and the additional flow-through streamlines making one horseshoe turn.
The angular momentum transferred to and from the planet by material on the horseshoe turns can be expressed in terms of the 
inverse vortensity $w$ in the corotation region, and so this is 
evaluated along two contours $C_1$ and $C_2$ just in front of and behind the two horseshoe turns, as shown schematically in Fig.~\ref{fig:integral_contours}.
Specifically, the expression given in equation~11 of \citet{2003ApJ...588..494M} for this horseshoe torque is
\begin{align}
\Gamma_{\rm hs} = \frac{3}{4} \rp \Omega_{\rm p}^3 \int_0^{x_s} \left[ w_{C_1}(x) -w_{C_2}(x)\right] x^2 dx\, , \label{eq:genhs}
\end{align}
{ where $\Omega_{\rm p}$ is the Keplerian angular velocity at the planet position}.
As the Lorentz force is a source of vortensity, this approximation requires that the magnetically-induced radial flow of gas is slow, as defined above.
We also demonstrate this more formally in Appendix~\ref{sec:vortsource}.
We can then follow the steps for simplifying this integral in the case of a low-mass planet as performed by
 \citet{2014MNRAS.444.2031P} (now in the case of a moving disc, instead of a moving planet), to arrive at
\begin{align}
\Gamma_{\rm hs} = 2\pi \left( 1-\frac{\wc (t)}{w(\rp)}\right) \Sigma_{\rm p} \rp ^2 x_s \Omega_{\rm p} (-v_r) \label{eq:gammahsvr}\, .
\end{align}
Here, $\wc(t)$ is the representative inverse vortensity of the well-mixed corotation region.
The difference between the integral of inverse vortensity between the $C_1$ contour sitting in front of the planet and the $C_2$ contour sitting behind the planet
is due to the narrowing asymmetry of the corotation region, and hence the width 
of the strip of flow with the disc background vortensity passing on flow-through streamlines is included.
Thus, although the prime issue is the asymmetry of the corotation region and the value of the characteristic vortensity of the libration region,
the flow-through streamlines contribute to the torque through the width of this strip.
Nondimensionalizing the torque from equation~(\ref{eq:gammahsvr}) 
by $\Gamma_0 = (q/h)^2 \Sigma_{\rm p} r_{\rm p}^4 \Omega_{\rm p}^2$,
and introducing the parameter $\chi$, yields
\begin{align}
\frac{\Gamma_{\rm hs}}{\Gamma_0} 
&= \frac{3}{2} \left( 1-\frac{\wc(t)}{w(\rp)}\right)  \left( \frac{q}{h} \right)^{-1/2} 
     \frac{   (1.2)^3  }{  \chi} \label{eq:gammahschi}
\end{align}
as an approximation for the horseshoe (corotation) torque on a low mass planet when $\chi\gg 1$.

To make use of this expression we then need to derive the evolution of the characteristic
 inverse vortensity $\wc$ of the material trapped in the libration region.
Here, we now finally consider explicitly that vortensity is not conserved in the disc when the Lorentz 
force due to the spiral magnetic field is present, but instead this torque causes $\wc$ to slowly evolve with time.
Starting from the continuity equation and the momentum equation in the (rotating) frame of the planet 
\begin{align}
\frac{\partial \Sigma}{\partial t} + \nabla\cdot(\Sigma \V)=0\\
\frac{\partial \V}{\partial t} +(\V\cdot\nabla)\V + 2\bm{\Omega}\times\V &= -\frac{1}{\Sigma}\nabla P - \nabla \Phi + \frac{T_\phi }{r \Sigma}\bm{\hat{\phi}} \\
T_\phi &= L_z r J_z \times B_r
\end{align}
{ where $P$ is the gas pressure, $\Phi$ is the gravitational potential, and $\bm{T}$ is the torque due to the magnetic field integrated over the height of the slab as in equation~(\ref{eq:vr})},
we form the evolution equation for absolute vortensity \citep[e.g.][p.~197]{salmon98}
\begin{align}
   \bm{\omega}_a &\equiv \nabla\times (\V+\bm{\Omega}\times\bm{r}) \\
\left(\frac{\partial }{\partial t} + \V\cdot \nabla\right)& \left(\frac{\bm{\omega}_a}{\Sigma}\right) 
    \nonumber\\
   & =\left[\frac{\bm{\omega}_a}{\Sigma}\cdot \nabla\right] \V +\frac{1}{\Sigma}\nabla P \times \nabla\frac{1}{\Sigma} +\frac{1}{\Sigma}\nabla\times\left(\frac{\bm{T}}{r\Sigma}\right)
\end{align}
where $\bm{\omega}_a$ is the absolute vorticity.
To apply this to our disc-planet system, we identify two regimes:
\begin{enumerate}
\item Outside closed streamlines and away from the planet, the flow is time-steady, and the radial drift of gas due to the Lorentz force ensures that 
\begin{align}
 \frac{\partial }{\partial t}  \left(\frac{\bm{\omega}_a}{\Sigma}\right) \approx 0
\end{align}
excepting the action of the weak shocks in the planet wake.
\item Inside closed streamlines, in a time averaged sense, 
\begin{align}
\frac{\partial }{\partial t}  \left(\frac{\bm{\omega}_a}{\Sigma}\right) 
   &\approx \frac{1}{\Sigma}\nabla\times\left(\frac{\bm{T}}{r\Sigma}\right) \label{eq:corotvort}
\end{align}
as $\bm{\omega}_a \cdot \nabla\V = 0$ due to the two-dimensionality of the flow 
and $\V\cdot\nabla(\bm{\omega}_a/\Sigma)\approx 0$ as the internal gradients in the 
well-mixed corotation region are small. 
For simplicity, we also assume that the flow is barotropic.
\end{enumerate}
Modelling the evolution of the inverse vortensity of the corotation region $\wc$ with equation~(\ref{eq:corotvort}) yields
\begin{align}
\frac{\partial \wc}{\partial t} &= - \wc^2 \left[ \frac{1}{\Sigma}\nabla\times\left(\frac{\bm{T}}{r\Sigma} \right)\right]
\end{align}
which can be solved as
\begin{align}
\wc(t) &= w(\rp) \left[ 1 + t/\tau_w \right]^{-1} \label{eq:wcsol}\\
\tau_w  &= \frac{4\pi \chi}{3(1.2)^2 \Omega_0} \left(\frac{3}{2}-\alpha \right)^{-1} \left(\frac{q}{h}\right)^{-1} \left(\frac{\rp}{r_0}\right)^{3/2},
\end{align}
using the inverse vortensity of the background disc model at the planet position $w(\rp)$ as the initial value and 
introducing the time-scale $\tau_w$ as the characteristic time for the vortensity of the material trapped in the corotation region to evolve.
We can then use this form in equation~(\ref{eq:gammahschi}) to calculate the corotation torque on the planet.
The total torque will be the sum of this corotation torque and the Lindblad (wave) torque.

Several aspects of these expressions should be noted:
\begin{enumerate}
\item${\Gamma_{\rm hs}}/{\Gamma_0}$ is positive for either sign of $\chi$ as the $1-\wc(t)/w(\rp)$ term changes sign with $\chi$. 
Thus for discs with $\alpha <3/2$ the change to the total torque is always positive, either slowing the inward migration or pushing the planet radially outwards.
\item Expanding equations~(\ref{eq:gammahschi}) and (\ref{eq:wcsol}) in powers of $t$ about zero shows the torque rises linearly with time at early times, with the same value for both signs of $\chi$.
\item The late time behaviour differs  for each sign of $\chi$. For positive $\chi$ in an $\alpha <3/2$ disc the 
inverse vortensity $\wc$ asymptotically decreases to zero, and the positive corotation torque 
asymptotes to a constant at late times. For negative $\chi$ in an $\alpha <3/2$ disc $\wc$ has a singularity at $t=\tau_w$
and the corotation torque is unbound. 
Thus, longer time evolution must be considered in the context of dynamical torques on live (self-consistently migrating) planets.
\item If we expand equations~(\ref{eq:gammahschi}) and (\ref{eq:wcsol}) in powers of $1/\chi$ about zero, 
the leading order term is of order $1/\chi^2$. 
Thus, in our numerical experiments we will scan the $\chi$ parameter by even steps in $1/\chi^2$.
\item The surface density $\Sigma_0$ of the disc scales out of the problem when appropriately nondimensionalized.
\end{enumerate}
It now remains to verify this expression in numerical simulations.

\section{Numerical Model}
\label{sec:numericalmodels}

\subsection{Methods}
\label{sec:numericalmethods}

For models with a live magnetic field, that is where we solve the induction equation as part of the simulation,
we use {\sc NIRVANA3.5} \citep{2004JCoPh.196..393Z}, which includes
Runge-Kutta-Legendre polynomial based second-order accurate
super-time-stepping \citep{2012MNRAS.422.2102M} of the Ohmic diffusion
operator, as previously employed in \citet{2015ApJ...801...84G},
with a newly implemented 
orbital advection scheme following \citet{2012A&A...545A.152M}.
Once we establish the equivalence of a live magnetic field, and an equivalent body force for the problem studied here
we revert to  {\sc FARGO3D} \citep{2015ascl.soft09006B,2016ApJS..223...11B} for the remainder of the simulations using 
purely hydrodynamic flow with a body force equivalent to the spiral magnetic field.

In both codes the boundary conditions employed are the same, combining fixed values at the initial condition on the boundary with damping zones in the style of
\citet{2006MNRAS.370..529D} driving all values towards the initial condition in an azimuthal ring close to the boundary, { for a variable $f$ with initial value $f_0$ this is}:
\begin{align}
\frac{d f}{dt} = -\frac{f-f_0}{\tau}\left[{\rm erf}(2x)\right]^4,
\end{align}
where ${\rm erf}$ is the error function and $x$ varies linearly between $1$ and $0$ from the edge of the domain to the inner edge of 
the damping zone, and the damping time-scale $\tau$ is $0.1$ local orbits.
These damping zones have the radial width $0.1$ at the inner boundary, and $0.2$ at the outer boundary.
As when $\chi \neq 0$ gas flows radially through the domain, the damping of the density 
field back to the initial condition provides the needed mass source and sink.
The background disc model derived in Section~\ref{sec:discmodel} is not in exact equilibrium. 
Hence, we allow the initial condition to relax for $10$ orbits before introducing the planet potential.

\subsection{Equivalence of live and fixed magnetic fields}
\label{sec:equiv}

\begin{figure}
\includegraphics[width=\columnwidth]{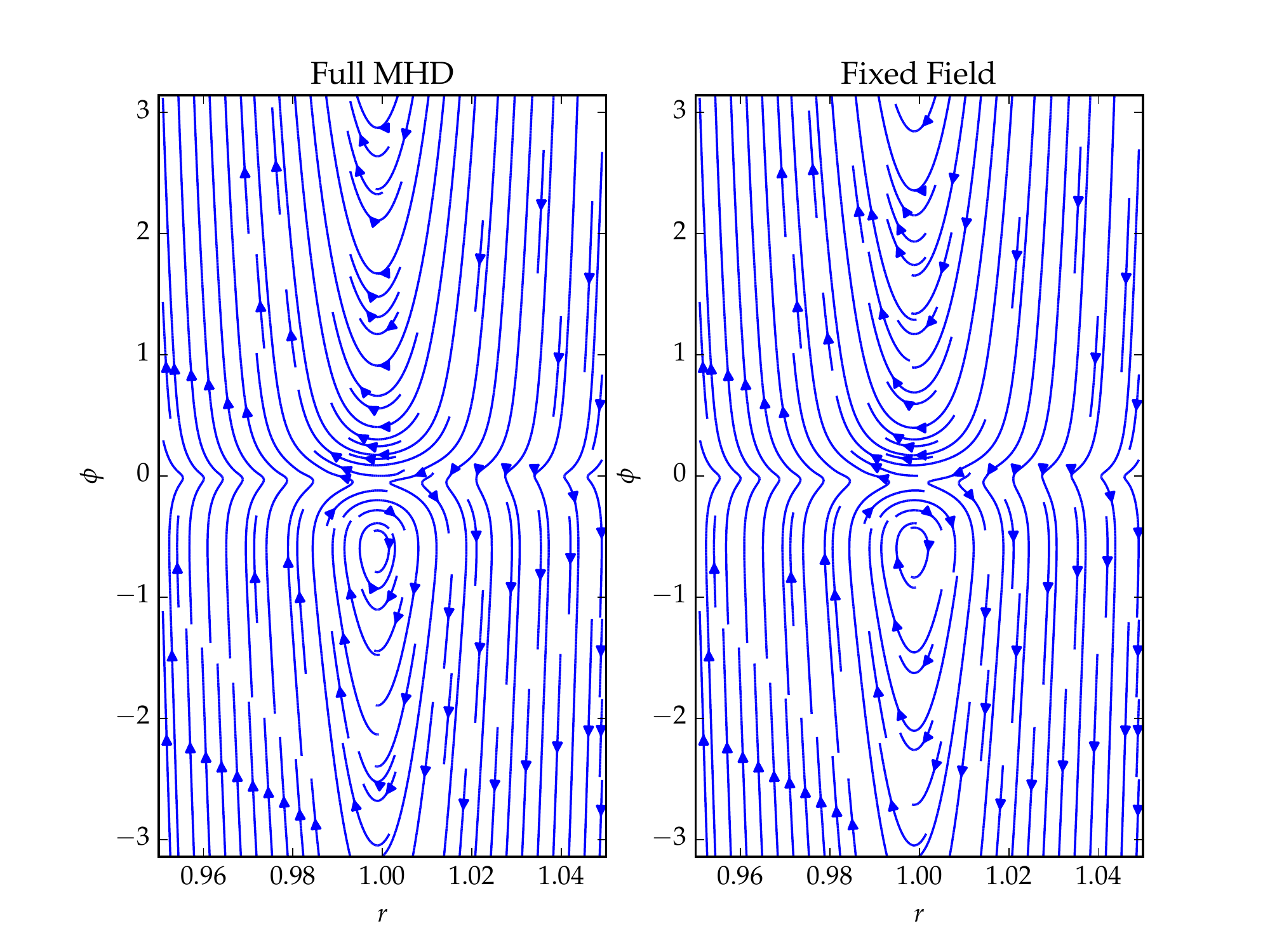}
\caption{{ Comparison of the flow streamlines in the coorbital region with a live magnetic field and a fixed magnetic field demonstrating that they are equivalent.
Left-hand panel: full live magnetic field. Right-hand panel: magnetic field fixed at the initial value.}}
\label{fig:sa6stream}
\end{figure}

\begin{figure}
\includegraphics[width=\columnwidth]{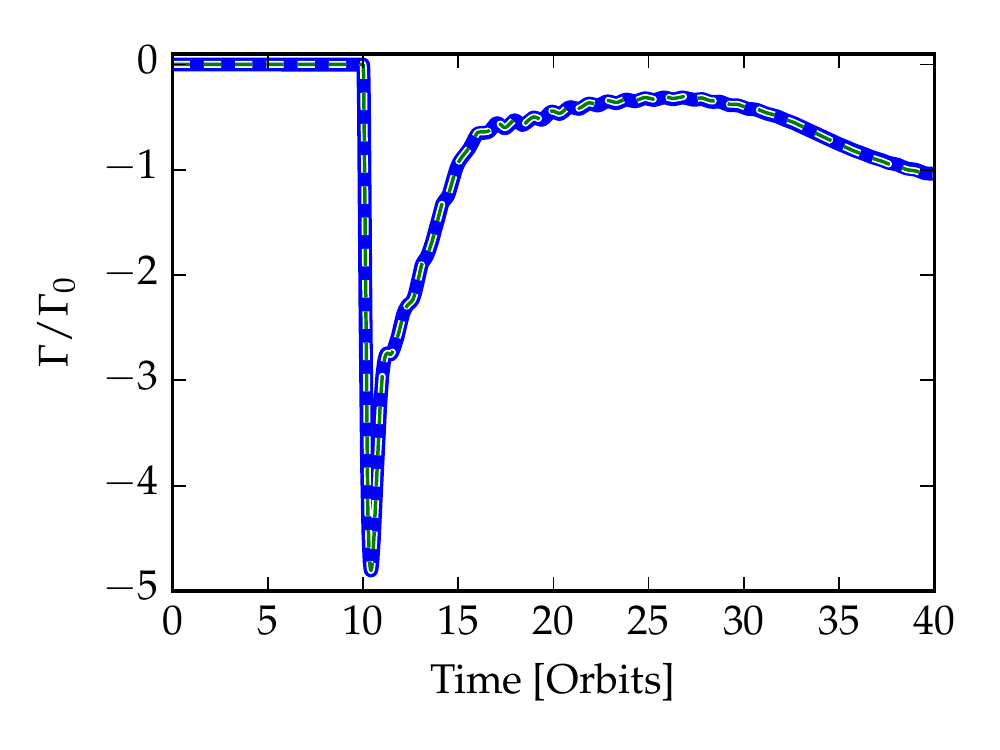}
\caption{Comparison between live magnetic field and equivalent hydrodynamic body force models demonstrating that they are equivalent.
Blue: full live magnetic field. Dashed green: hydrodynamic model with an equivalent body force.}
\label{fig:sa6}
\end{figure}

To demonstrate the equivalence of live magnetic field, modelled with the induction equation in full resistive magnetohydrodynamics,
and an equivalent body force to the initial condition field equations~(\ref{eq:bric}) and~(\ref{eq:bphiic}),
we run a single model with a low $\chi$ value.
To keep the disc model stable with a live magnetic field, we found it necessary to choose parameters
such the the plasma beta at the inner boundary remains high.
As such, we employ a surface density profile $\Sigma = r^{0}$ and a locally isothermal temperature profile $c_s^2=h= (0.05)^2 r^{-2}$.
A planet mass ratio $q=2\times10^{-5}$ is used to maximize the width of the corotation region, with smoothing of $r_s=0.4h$.
The grid extends in radius $[0.19619,1.767]$ with radial resolution $512$ cells and azimuthal $1536$ cells both evenly spaced.
The magnetic field has a plasma beta at the planet position of $\beta_0=10$ and $\chi=1$, this produces a significant flow past the planet position.
{ We show comparisons in two stages. 
First, we compare the the flow pattern in the coorbital region for a full MHD calculation with a live magnetic field, 
and a calculation with the magnetic field fixed to the initial condition value, in Fig.~\ref{fig:sa6stream}, at $t=40$~orbits.
These can be seen to be very similar.
Then, we replace the fixed magnetic field with a body force exerting the same Lorentz force as the fixed magnetic field and show the
resulting planet torque history is shown in Fig.~\ref{fig:sa6}. }
Once the planet is introduced at 10 orbits, 
the torque evolution is nearly indistinguishable.
Thus, for the rest of this paper we continue our experiments using only purely hydrodynamic models with {\bf such} an applied body force.

\subsection{Torques with an equivalent body force}

\begin{figure}
\includegraphics[width=\columnwidth]{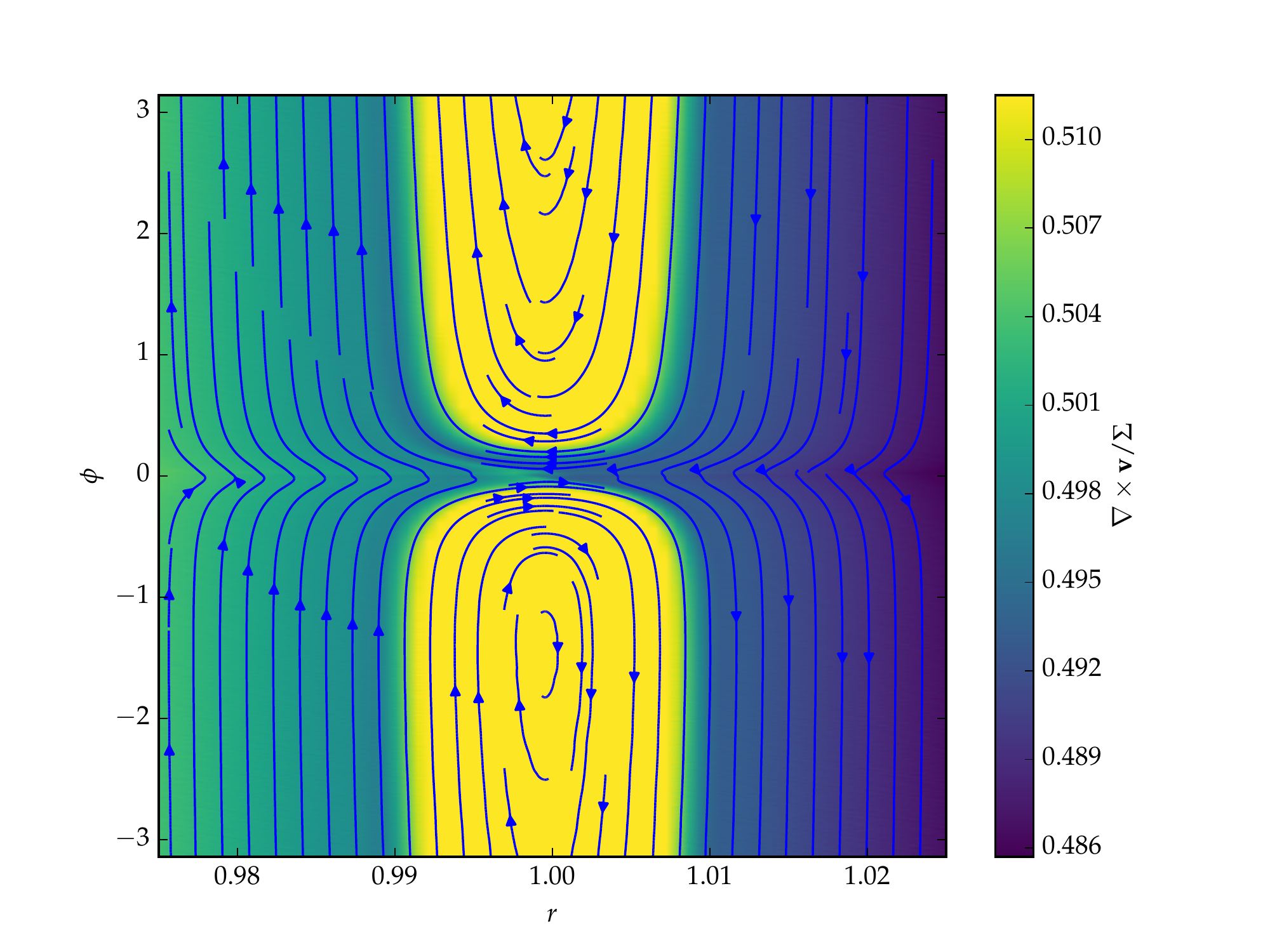}
\caption{Flow in the corotation region for $\chi=5\sqrt{2}$ at $T=1000$ orbits with streamlines of velocity in the frame 
of the planet drawn in blue, and absolute vortensity shown in the colour map.
Note both the high vortensity of the libration region, and the asymmetrical shape.}
\label{fig:corotflow}
\end{figure}

\begin{figure}
\includegraphics[width=\columnwidth]{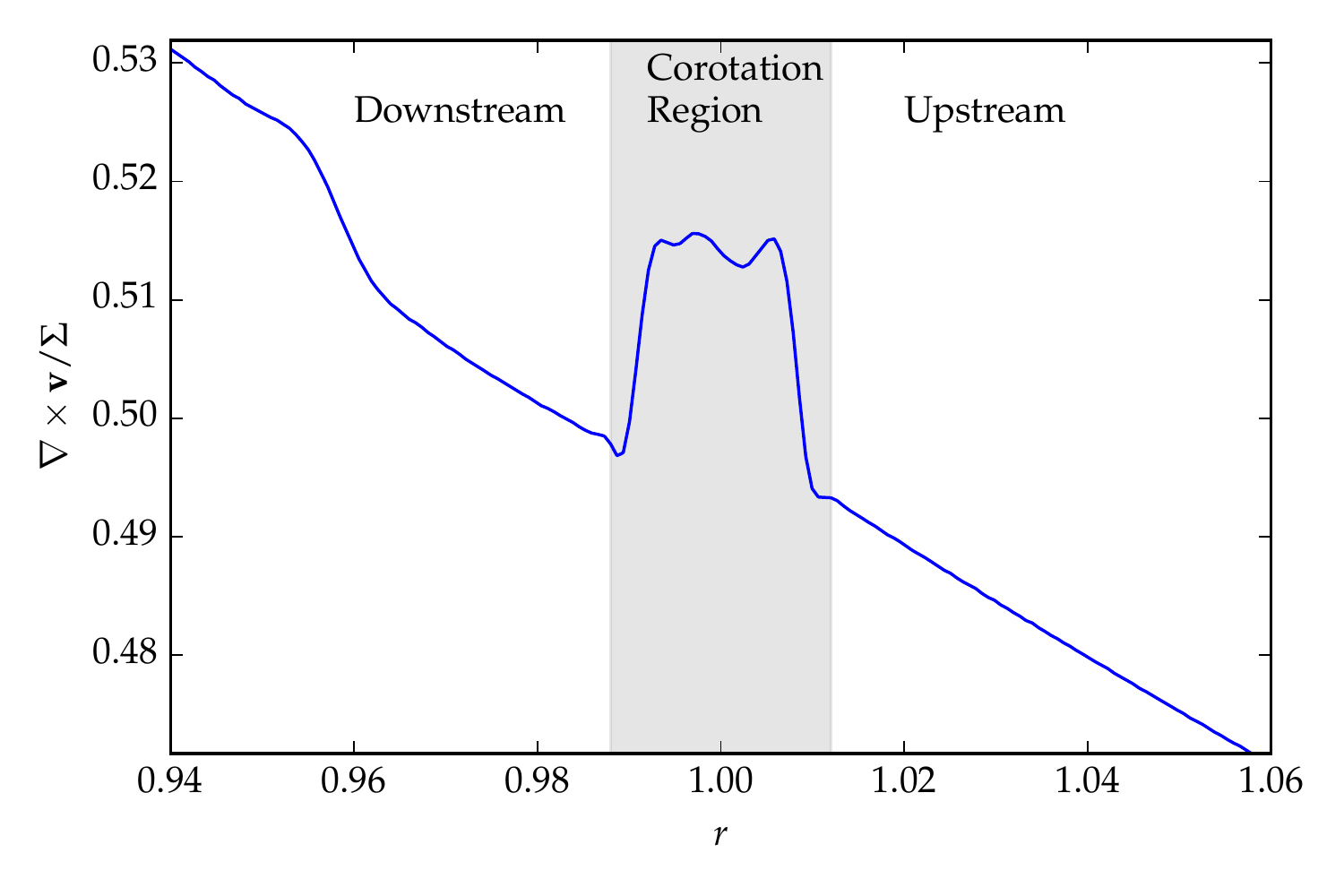}
\caption{Cut across the corotation region for $\chi=\sqrt{2}$, opposite of the planet position, showing vortensity at $T=1000$ orbits.
The vortensity on the librating streamlines in the corotation region has been driven by the Lorentz force, 
while upstream the disc remains at its initial vortensity, and downstream the material which has passed by the planet 
on flow-through streamlines has continued to spiral towards the star. At approximately $r=0.96$ the transition between material which has passed the planet once, and material which was interior to the planet at the beginning of the simulation is evident.}
\label{fig:cuts}
\end{figure}

\begin{figure}
\includegraphics[width=\columnwidth]{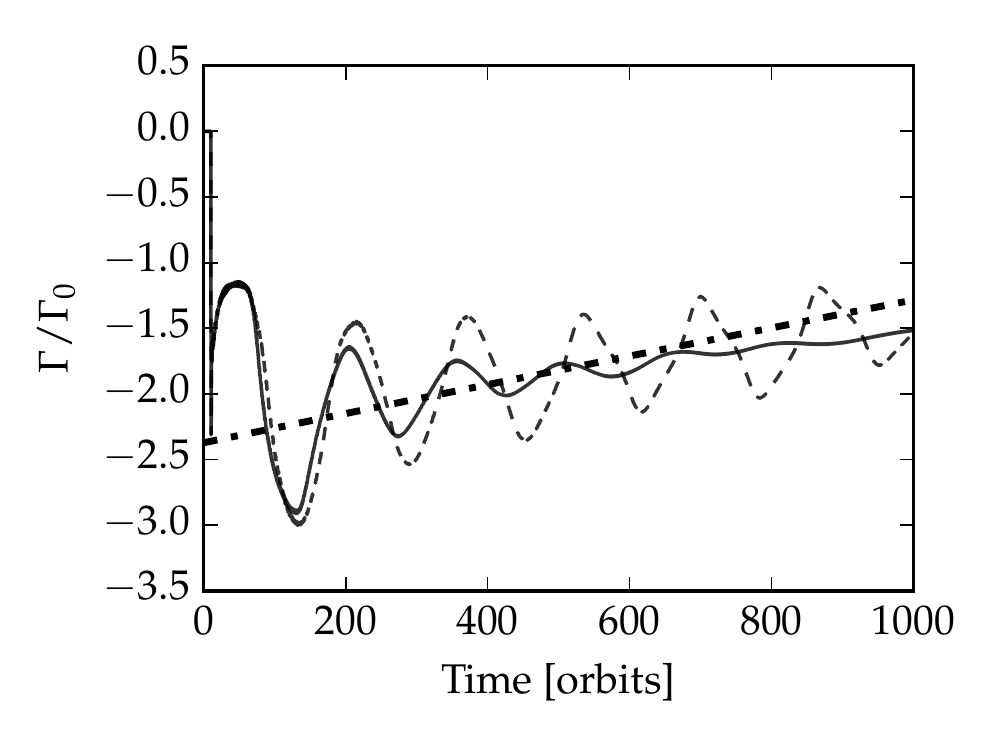}
\includegraphics[width=\columnwidth]{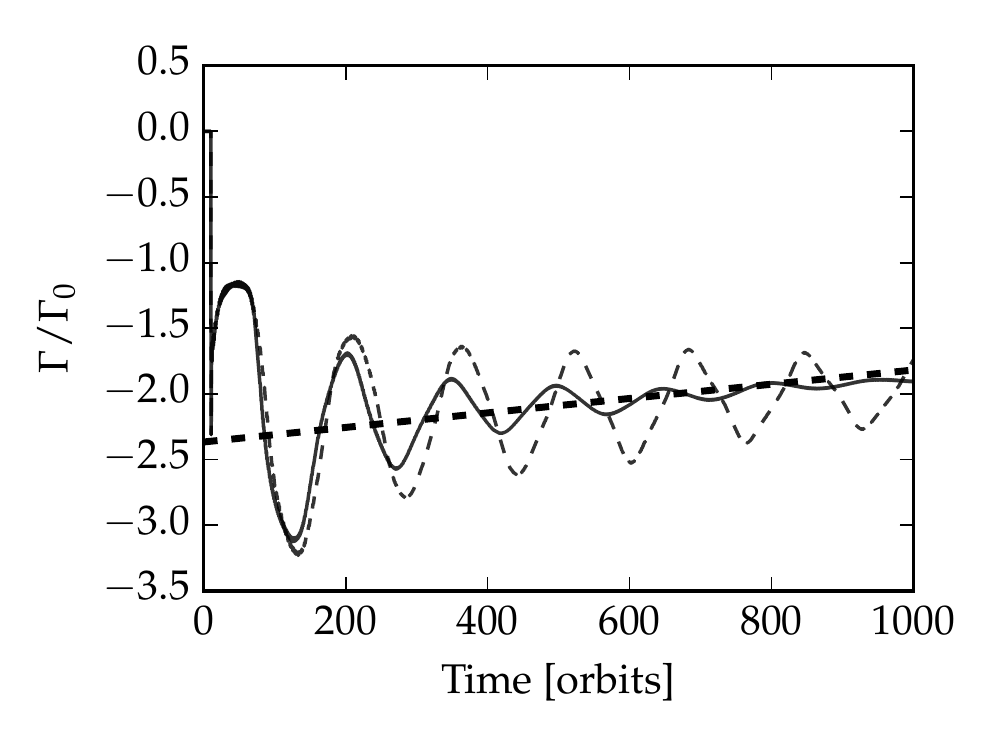}
\includegraphics[width=\columnwidth]{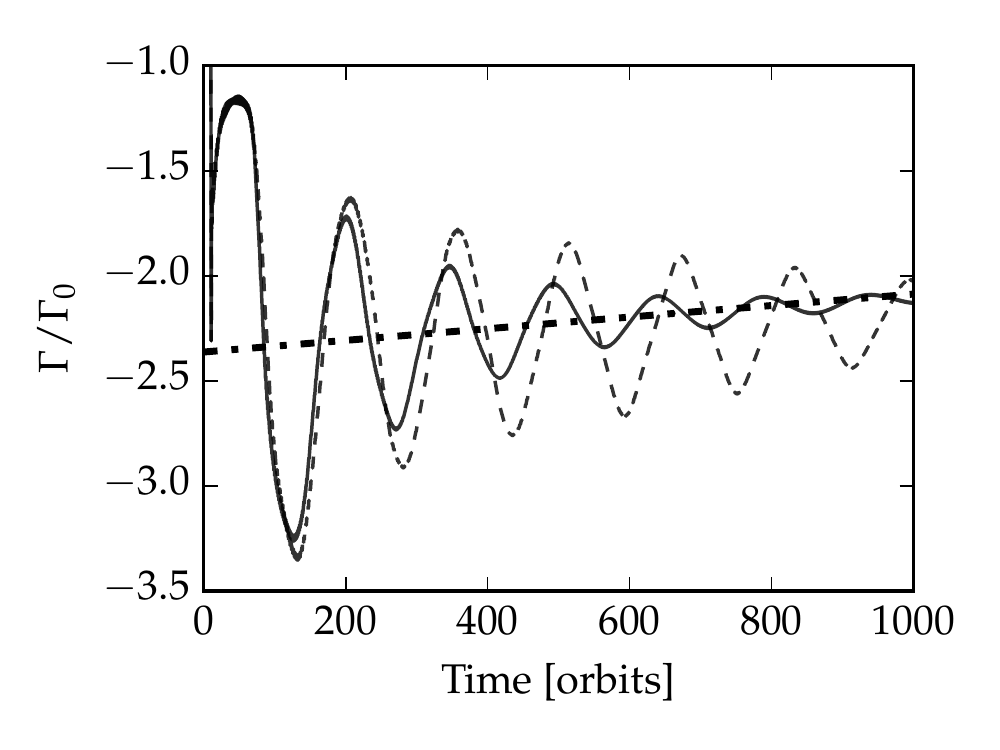}
\caption{Torque histories for low mass planets, scan of $\chi$ parameter.
Top panel: $\chi=\pm 5\sqrt{2}$, Middle panel: $\chi=\pm 10$, Bottom panel: $\chi = \pm 10\sqrt{2}$
Solid lines: positive $\chi$. Dashed lines: negative $\chi$. Dot--dashed lines:  Analytical curve predicted from equation~(\ref{eq:gammahschi}), with an arbitrary position on the vertical axis.}
\label{fig:positivechiscan}
\end{figure}

Having established that the magnetic field does not locally respond to the flow around the low mass planet, 
we proceed with models  tuned to testing the torque formula presented in Section~\ref{sec:torqueformula}.
These are purely hydrodynamic models that use only a body force given as the Lorentz force produced by the magnetic field
equations~(\ref{eq:bric})-(\ref{eq:bphiic}).
We employ {\sc FARGO3D}, on the same domain as in the previous section with resolution 
$N_r, N_\phi =[2048,6144]$
and a lower mass planet with
$q=5\times 10^{-6}$ corresponding to a planet of $1.7\ {\rm M_\oplus}$ orbiting a solar mass star.
A shallow density profile of
$\Sigma =  r^{-1/2}$ is set and the disc is globally isothermal with $c_s^2=0.05$
so that the only vortensity source is provided by the body force.
We scan $\chi$ as $-10\sqrt{2}$, $-10$, $-5\sqrt{2}$, $5\sqrt{2}$, $10$, and $10\sqrt{2}$.
To numerically realize the flow with negative values of $\chi$, we use the equivalent body force which would be produced with 
a negative value of $\eta$ in equation~(\ref{eq:bphiic}).

Aside from the spiral wake, the planet barely affects the density distribution of the disc, but the trapped material on
 librating streamlines in the corotation region gradually evolves.
For the positive $\chi$ values, the vortensity increases with time and the libration region has an asymmetrical shape.
This is shown in Fig.~\ref{fig:corotflow}, where for the $\chi=5\sqrt{2}$ model the vortensity 
and streamlines are shown after 1000 orbits.
In more detail, a cut opposite the planet position, or equivalently at the bottom of 
Fig.~\ref{fig:corotflow}, is shown in Fig.~\ref{fig:cuts}.
This shows in detail the well-mixed top-hat like distribution of vortensity in the corotation region, 
and how the surrounding disc flows past the planet unmodified, simply shifted in by the 
single pass near the planet on the flow-through streamlines.
Here it is evident again how this effect differs from viscous desaturation of the corotation torque - 
the sharp edges in vortensity here are only preserved because of the very low numerical viscosity.

Comparisons of the evolution of the total torque in this suite of simulations 
with the theory given by equations~(\ref{eq:gammahschi}) and (\ref{eq:wcsol}) 
are presented in Fig.~\ref{fig:positivechiscan}.
The $\chi=10\sqrt{2}$ case is the subject of a resolution study presented in Appendix~\ref{sec:resolutionstudy}.
High resolution is needed in the corotation region to preserve the growth of the vortensity on librating streamlines.

\section{Discussion}
\label{sec:discussion}

Our reduced model of the dead zone assumes that the corotation region is relatively unaffected
by the action of the wind and accreting surface layers of the disc. 
Although in isothermal hydrodynamic models it is found that the fluid in the corotation region rotates in the same sense at all heights above the mid plane \citep{2016ApJ...817...19M},
the much higher densities in the dead zone mean that the low-density flow 
in the active one cannot exert a large Reynolds stress at the vertical active-dead 
(or more precisely magnetically coupled-uncoupled) interface.
Precisely verifying this will require running full three dimensional magnetohydrodynamic
models explicitly including Ohmic, ambipolar, and Hall terms, 
and sufficiently self-consistent ionization chemistry.

Hall-stress enabled corotation torques differ from corotation torques that arise because of viscous desaturation. Fundamentally, the difference occurs because in a very low or zero viscosity situation, the corotation region has a very long memory of the vortensity sources applied to it. Viscous desaturation of the corotation torque relies precisely on removing this long memory, 
by mixing the vortensity of the corotation region with that of the surrounding disc 
across the flow separatrix at the edge of the libration island. We present the following explanation to clarify how the corotation torque on a low mass planet differs in viscous discs versus laminar discs with magnetic torques.
\begin{enumerate}
\item Ignoring radial motion of the planet or disc material, a sustained corotation torque in an isothermal disc arises when a vortensity gradient exists across the horseshoe region. In the absence of viscosity, any vortensity gradient present initially across the horseshoe region is removed through phase mixing by the horseshoe motion, saturating (switching off) the corotation torque. Viscosity is required to maintain the vortensity gradient in the disc and prevent the saturation of the corotation torque, which it does by diffusing vortensity across the horseshoe region and across the boundary between the horseshoe region and the surrounding disc. In a steady disc, this corotation torque is independent of time and depends only on the local disc and planet parameters. This is the viscously desaturated corotation torque that has been considered in numerous earlier studies \citep[e.g.][]{2009MNRAS.394.2283P}.

\item When radial motion of the disc material due to a magnetic torque and/or radial migration of the planet is considered in an inviscid isothermal disc, then a corotation torque can arise because of the existence of flow-through streamlines, distortion of the geometry of the horseshoe region and evolution of the ratio of the inverse vortensity in the corotation region to that in the surrounding disc, $w_{\rm c}/w_{\rm d}$. Considering the case of a non-migrating planet and a magnetically torqued disc that leads to inflow of the gas at a spatially and temporally constant rate, the magnetic torque will cause $w_{\rm c}/w_{\rm d}$ to evolve with time, leading to continuous time evolution of the corotation torque as expressed by equation~(\ref{eq:gammahschi}). The absence of viscosity prevents any mixing of vortensity between the corotation region and the surrounding disc, ensuring that $w_{\rm c}$ and $w_{\rm d}$ evolve independently of one another, and hence there is no steady corotation torque in this case. Furthermore, at any point in time the value of $w_{\rm c}/w_{\rm d}$, and hence the corotation torque, depends on the integrated history of the magnetic torques that have been applied to the corotation region, such that it retains memory of the initial conditions and past evolution.

\item We can now consider the case of a steadily accreting viscous disc in which the surface density and radial velocity profiles are identical to those in the laminar, magnetically torqued disc discussed above \citep{2001ApJ...558..453M}. The net torque acting at each radius in this hypothetical viscous disc is identical to that in the magnetically torqued disc, and so the arguments leading to the development of equation~(\ref{eq:gammahschi}) apply equally to this case. The fundamental difference between the viscous and non-viscous disc, however, is that angular momentum and hence vortensity are transported diffusively in the viscous disc, such that any evolution of the ratio $w_{\rm c}/w_{\rm d}$ that might arise because of local torques acting is counterbalanced by diffusive mixing of the vortensity between the corotation region and the surrounding disc. This diffusion continuously drives $w_{\rm c}/w_{\rm d} \rightarrow 1$, such that the dynamical corotation torque given by equation~(\ref{eq:gammahschi}) is equal to zero. The corotation torque operating in this case is then just the viscously desaturated version described above. This discussion indicates that if viscosity was introduced into our simulations with a non-migrating planet, and increased such that it 
dominates over the magnetic torque, the corotation torque would saturate to a steady value.

\end{enumerate}

The corotation torque effect in a laminar magnetically torqued disc, or moreover the effect on the evolution of the vortensity of the corotation region
due to Hall-effect driven magnetic fields, is fundamentally related to the dynamical corotation torque experience by a migrating planet as presented by
\citet{2014MNRAS.444.2031P}.
Both are in the end due to the relative radial movement of the planet and disc material.
In terms of the corotation torque, negative $\chi$ values are equivalent to inward migration of the planet,
and positive $\chi$ values are equivalent to outward migration of the planet.
Thus, we can consider what the runaway inward or outward migration, or 
stopping scenarios are under the joint action of corotation and Lindblad torques in a disc with a radial flow induced by a magnetic torque. 
The appropriate form for a unified torque is
\begin{align}
\Gamma_{\rm hs} = 2\pi \left( 1-\frac{\wc (t)}{w(\rp)}\right) \Sigma_{\rm p} \rp ^2 x_s \Omega_{\rm p} \left[\frac{d \rp}{dt} -v_r \right]
\label{eq:unified}
\end{align}
where the response of the planet changing its migration rate 
will prevent the torque from growing without bound.

There are three different discriminants that determine how equation~(\ref{eq:unified}) behaves;
the sign of $\chi$ (equivalently  $-v_r$), the sign of $d \rp/dt$ and the sign of $[d \rp/dt-v_r]$.
Here we discuss the situation of a  disc with a surface density slope that is shallower than $\alpha=3/2$, 
and suppose  that the planet is initially migrating inwards, presumably under the influence of Lindblad torques. Before discussing what happens for values of $\chi \ne 0$, we first recall what happens for a planet migrating in an inviscid disc without any magnetic torque acting.

The vortensity profile in a quasi-Keplerian disc scales according to $(\nabla \times v)/\Sigma \sim \Omega/\Sigma \sim (\Omega_0/\Sigma_0) r^{\alpha-3/2}$. Hence, the inverse vortensity scales as $w \sim r^{3/2-\alpha}$, such that $w$ decreases as one moves closer to the star. When no magnetic torque acts, $v_r=0$ in equation~(\ref{eq:unified}), and as the planet migrates inwards (preserving the inverse vortensity of the material in the corotation region, $\wc$) the ratio $\wc/w(\rp)$ increases above unity and the corotation torque becomes increasingly positive, slowing the inwards migration.
Now we consider what happens when $\chi \ne 0$.
\begin{enumerate}
\item For positive values of $\chi$, that is inwards moving disc gas, 
if the planet is moving inwards faster than the disc gas, then $[d \rp/dt-v_r] <0$, and the evolution will initially be similar to the case described above with $\chi=0$. The corotation torque will increase as the planet migrates inwards, but the magnetic torque will act to slowly decrease $\wc(t)$ in equation~(\ref{eq:unified}), such that the corotation torque will not increase as quickly as it would with $\chi=0$. Nonetheless, the increasing corotation torque slows the planet migration until $d\rp/dt = v_r$, after which the corotation torque switches off and the planet migrates inwards at the same speed as the inflowing gas.

\item For positive values of $\chi$ and $|v_r| > |d\rp/dt|$ (i.e. disc gas moving inwards more quickly than the planet), we have $[d \rp/dt-v_r] >0$. Now the magnetic torque causes $\wc$ to decrease quickly relative to $w(\rp)$, the first term in parentheses in equation~(\ref{eq:unified}) is positive and increasing, giving a positive and increasing corotation torque. The inwards planet migration can thus be halted and reversed, leading to runaway outwards migration because the corotation torque increases as $d\rp/dt$ increases.

\item For negative values of $\chi$ the disc gas is moving outwards, and for a planet initially migrating inwards $[d \rp/dt-v_r] <0$. The inwards migration causes $\wc/w(\rp)$ to increase above unity, and the magnetic torque also acts to increase $\wc$. Hence, the corotation torque is positive and increasing such that the migration can be stopped and reversed. The outward migration speed of the planet increases until $d\rp/dt = v_r$, and the corotation torque then switches off, leaving the planet to migrate along with the outwards flowing gas.

\item If $\chi <0$, and for some reason the planet is initially able to migrate outwards faster than the disc gas, then $[d \rp/dt-v_r] > 0$. Because of the background inverse vortensity profile, outwards migration of the planet causes the first term in parentheses in equation~(\ref{eq:unified}) to increase towards unity, giving a positive corotation torque that can cause a runaway because $\Gamma_{\rm hs}$ scales with $d \rp/dt$.

\end{enumerate}

We note that if the slope of the surface density is steeper than $\alpha=3/2$, then 
the background disc vortensity increases outwards.
The dynamical corotation torque on its own now exhibits a positive feedback on migration \citep{2014MNRAS.444.2031P}.
Likewise, the combined torque in equation~(\ref{eq:unified}) results in a feedback which 
always drives planets inwards.
Finally, our analysis in this work has been limited to barotropic discs, but the extension of dynamical torques to 
radiative baroclinic, and viscously accreting discs has been considered in  \citet{2015MNRAS.454.2003P} and \citet{2016MNRAS.462.4130P} where an 
additional corotation torque effect due to a entropy contrast dynamically built up in the librating streamlines has been shown.
Likewise, in a baroclinic disc with laminar magnetic torques we expect additional thermodynamic contributions to the corotation torque to arise.

\section{Conclusions}
\label{sec:conclusions}

We have demonstrated that a midplane laminar Hall-enabled Maxwell stress that drives radial flow 
(either radially inwards or outwards) results in a sustained and growing corotation torque on low mass planets.
For a planet held on a fixed orbit in a globally isothermal disc, we can accurately account for this torque 
in terms of the vortensity source due to the action of the magnetic field on the material trapped in the libration region.
Furthermore, we have proposed that this effect due to the radial motion of the disc gas past the planet and the dynamical corotation torque
due to the radial motion of the planet past the disc can be unified into a single treatment.
We have suggested that when laminar Maxwell stress drives accretion in the midplane, 
the radial migration of low mass planets can be slowed or reversed, and possibly run away outwards even in an inviscid disc. These predicted behaviours will be tested in a study of planets migrating in discs that sustain radial flows in their midplanes, to be presented in a  forthcoming paper.

One overarching lesson of this work and \citet{2014MNRAS.444.2031P} is that in low viscosity environments, 
the corotation torque is not determined instantaneously by local disc conditions. 
Instead, it reflects a significant hysteresis, or memory effect, of the history of the forcing of the corotation region.
This behaviour cannot be captured using an instantaneous torque formula as commonly applied to lower dimensional 
models of planet migration in viscous discs \citep{2010MNRAS.401.1950P,2011MNRAS.410..293P}.

\section*{Acknowledgements}

This research was supported by an STFC Consolidated grant awarded to the QMUL Astronomy Unit 2012-2016.
The simulations presented in this paper utilized Queen Mary's MidPlus computational facilities, supported by QMUL Research-IT and funded by EPSRC grant EP/K000128/1; and the DiRAC Complexity system, operated by the University of Leicester IT Services, which forms part of the STFC DiRAC HPC Facility (www.dirac.ac.uk). 
This equipment is funded by BIS National E-Infrastructure capital grant ST/K000373/1 and STFC DiRAC Operations grant ST/K0003259/1. DiRAC is part of the National E-Infrastructure.
This work used the NIRVANA3.5 code developed by Udo Ziegler at the
Leibniz Institute for Astrophysics (AIP).
This research was supported in part by the National Science Foundation under Grant No. NSF PHY11-25915. 
The research leading to these results has received funding from the European Research Council (ERC) under the European Union's Horizon 2020 research and innovation programme (grant agreement No 638596).
This research was supported by the Munich Institute for Astro- and Particle Physics (MIAPP) of the DFG cluster of excellence ``Origin and Structure of the Universe''. CPM, RPM, and OG thank the MIAPP for hospitality.  RPN and CPM thank the OCA for hospitality. SJP is supported by a Royal Society University Research Fellowship.




\bibliographystyle{mnras}

\begin{thebibliography}{}
\makeatletter
\relax
\def\mn@urlcharsother{\let\do\@makeother \do\$\do\&\do\#\do\^\do\_\do\%\do\~}
\def\mn@doi{\begingroup\mn@urlcharsother \@ifnextchar [ {\mn@doi@}
  {\mn@doi@[]}}
\def\mn@doi@[#1]#2{\def\@tempa{#1}\ifx\@tempa\@empty \href
  {http://dx.doi.org/#2} {doi:#2}\else \href {http://dx.doi.org/#2} {#1}\fi
  \endgroup}
\def\mn@eprint#1#2{\mn@eprint@#1:#2::\@nil}
\def\mn@eprint@arXiv#1{\href {http://arxiv.org/abs/#1} {{\tt arXiv:#1}}}
\def\mn@eprint@dblp#1{\href {http://dblp.uni-trier.de/rec/bibtex/#1.xml}
  {dblp:#1}}
\def\mn@eprint@#1:#2:#3:#4\@nil{\def\@tempa {#1}\def\@tempb {#2}\def\@tempc
  {#3}\ifx \@tempc \@empty \let \@tempc \@tempb \let \@tempb \@tempa \fi \ifx
  \@tempb \@empty \def\@tempb {arXiv}\fi \@ifundefined
  {mn@eprint@\@tempb}{\@tempb:\@tempc}{\expandafter \expandafter \csname
  mn@eprint@\@tempb\endcsname \expandafter{\@tempc}}}

\bibitem[\protect\citeauthoryear{{Bai}}{{Bai}}{2013}]{2013ApJ...772...96B}
{Bai} X.-N.,  2013, \mn@doi [\apj] {10.1088/0004-637X/772/2/96}, \href
  {http://adsabs.harvard.edu/abs/2013ApJ...772...96B} {772, 96}

\bibitem[\protect\citeauthoryear{{Bai}}{{Bai}}{2014a}]{2014ApJ...791...73B}
{Bai} X.-N.,  2014a, \mn@doi [\apj] {10.1088/0004-637X/791/1/73}, \href
  {http://adsabs.harvard.edu/abs/2014ApJ...791...73B} {791, 73}

\bibitem[\protect\citeauthoryear{{Bai}}{{Bai}}{2014b}]{2014ApJ...791..137B}
{Bai} X.-N.,  2014b, \mn@doi [\apj] {10.1088/0004-637X/791/2/137}, \href
  {http://adsabs.harvard.edu/abs/2014ApJ...791..137B} {791, 137}

\bibitem[\protect\citeauthoryear{{Bai}}{{Bai}}{2015}]{2015ApJ...798...84B}
{Bai} X.-N.,  2015, \mn@doi [\apj] {10.1088/0004-637X/798/2/84}, \href
  {http://adsabs.harvard.edu/abs/2015ApJ...798...84B} {798, 84}

\bibitem[\protect\citeauthoryear{{Bai}}{{Bai}}{2016}]{2016ApJ...821...80B}
{Bai} X.-N.,  2016, \mn@doi [\apj] {10.3847/0004-637X/821/2/80}, \href
  {http://adsabs.harvard.edu/abs/2016ApJ...821...80B} {821, 80}

\bibitem[\protect\citeauthoryear{{Bai} \& {Stone}}{{Bai} \&
  {Stone}}{2013}]{2013ApJ...769...76B}
{Bai} X.-N.,  {Stone} J.~M.,  2013, \mn@doi [\apj]
  {10.1088/0004-637X/769/1/76}, \href
  {http://adsabs.harvard.edu/abs/2013ApJ...769...76B} {769, 76}

\bibitem[\protect\citeauthoryear{{Bai}, {Ye}, {Goodman}  \& {Yuan}}{{Bai}
  et~al.}{2016}]{2016ApJ...818..152B}
{Bai} X.-N.,  {Ye} J.,  {Goodman} J.,   {Yuan} F.,  2016, \mn@doi [\apj]
  {10.3847/0004-637X/818/2/152}, \href
  {http://adsabs.harvard.edu/abs/2016ApJ...818..152B} {818, 152}

\bibitem[\protect\citeauthoryear{{Baruteau} \& {Masset}}{{Baruteau} \&
  {Masset}}{2008}]{2008ApJ...672.1054B}
{Baruteau} C.,  {Masset} F.,  2008, \mn@doi [\apj] {10.1086/523667}, \href
  {http://adsabs.harvard.edu/abs/2008ApJ...672.1054B} {672, 1054}

\bibitem[\protect\citeauthoryear{{Baruteau} et~al.,}{{Baruteau}
  et~al.}{2014}]{2014prpl.conf..667B}
{Baruteau} C.,  et~al., 2014, \mn@doi [Protostars and Planets VI]
  {10.2458/azu_uapress_9780816531240-ch029}, \href
  {http://adsabs.harvard.edu/abs/2014prpl.conf..667B} {pp 667--689}

\bibitem[\protect\citeauthoryear{{Ben{\'{\i}}tez Llambay} \&
  {Masset}}{{Ben{\'{\i}}tez Llambay} \& {Masset}}{2015}]{2015ascl.soft09006B}
{Ben{\'{\i}}tez Llambay} P.,  {Masset} F.,  2015, {FARGO3D:
  Hydrodynamics/magnetohydrodynamics code}, Astrophysics Source Code Library
  (\mn@eprint {ascl} {1509.006})

\bibitem[\protect\citeauthoryear{{Ben{\'{\i}}tez-Llambay} \&
  {Masset}}{{Ben{\'{\i}}tez-Llambay} \& {Masset}}{2016}]{2016ApJS..223...11B}
{Ben{\'{\i}}tez-Llambay} P.,  {Masset} F.~S.,  2016, \mn@doi [\apjs]
  {10.3847/0067-0049/223/1/11}, \href
  {http://adsabs.harvard.edu/abs/2016ApJS..223...11B} {223, 11}

\bibitem[\protect\citeauthoryear{{B{\'e}thune}, {Lesur}  \&
  {Ferreira}}{{B{\'e}thune} et~al.}{2017}]{2017A&A...600A..75B}
{B{\'e}thune} W.,  {Lesur} G.,   {Ferreira} J.,  2017, \mn@doi [\aap]
  {10.1051/0004-6361/201630056}, \href
  {http://adsabs.harvard.edu/abs/2017A%26A...600A..75B} {600, A75}

\bibitem[\protect\citeauthoryear{{Fricke}}{{Fricke}}{1968}]{1968ZA.....68..317F}
{Fricke} K.,  1968, \zap, \href
  {http://adsabs.harvard.edu/abs/1968ZA.....68..317F} {68, 317}

\bibitem[\protect\citeauthoryear{{Fromang}, {Terquem}  \& {Nelson}}{{Fromang}
  et~al.}{2005}]{2005MNRAS.363..943F}
{Fromang} S.,  {Terquem} C.,   {Nelson} R.~P.,  2005, \mn@doi [\mnras]
  {10.1111/j.1365-2966.2005.09498.x}, \href
  {http://adsabs.harvard.edu/abs/2005MNRAS.363..943F} {363, 943}

\bibitem[\protect\citeauthoryear{{Goldreich} \& {Schubert}}{{Goldreich} \&
  {Schubert}}{1967}]{1967ApJ...150..571G}
{Goldreich} P.,  {Schubert} G.,  1967, \mn@doi [\apj] {10.1086/149360}, \href
  {http://adsabs.harvard.edu/abs/1967ApJ...150..571G} {150, 571}

\bibitem[\protect\citeauthoryear{{Goldreich} \& {Tremaine}}{{Goldreich} \&
  {Tremaine}}{1980}]{1980ApJ...241..425G}
{Goldreich} P.,  {Tremaine} S.,  1980, \mn@doi [\apj] {10.1086/158356}, \href
  {http://adsabs.harvard.edu/abs/1980ApJ...241..425G} {241, 425}

\bibitem[\protect\citeauthoryear{{Gressel}, {Nelson}  \& {Turner}}{{Gressel}
  et~al.}{2011}]{2011MNRAS.415.3291G}
{Gressel} O.,  {Nelson} R.~P.,   {Turner} N.~J.,  2011, \mn@doi [\mnras]
  {10.1111/j.1365-2966.2011.18944.x}, \href
  {http://adsabs.harvard.edu/abs/2011MNRAS.415.3291G} {415, 3291}

\bibitem[\protect\citeauthoryear{{Gressel}, {Nelson}  \& {Turner}}{{Gressel}
  et~al.}{2012}]{2012MNRAS.422.1140G}
{Gressel} O.,  {Nelson} R.~P.,   {Turner} N.~J.,  2012, \mn@doi [\mnras]
  {10.1111/j.1365-2966.2012.20701.x}, \href
  {http://adsabs.harvard.edu/abs/2012MNRAS.422.1140G} {422, 1140}

\bibitem[\protect\citeauthoryear{{Gressel}, {Nelson}, {Turner}  \&
  {Ziegler}}{{Gressel} et~al.}{2013}]{2013ApJ...779...59G}
{Gressel} O.,  {Nelson} R.~P.,  {Turner} N.~J.,   {Ziegler} U.,  2013, \mn@doi
  [\apj] {10.1088/0004-637X/779/1/59}, \href
  {http://adsabs.harvard.edu/abs/2013ApJ...779...59G} {779, 59}

\bibitem[\protect\citeauthoryear{{Gressel}, {Turner}, {Nelson}  \&
  {McNally}}{{Gressel} et~al.}{2015}]{2015ApJ...801...84G}
{Gressel} O.,  {Turner} N.~J.,  {Nelson} R.~P.,   {McNally} C.~P.,  2015,
  \mn@doi [\apj] {10.1088/0004-637X/801/2/84}, \href
  {http://adsabs.harvard.edu/abs/2015ApJ...801...84G} {801, 84}

\bibitem[\protect\citeauthoryear{{Guilet}, {Baruteau}  \&
  {Papaloizou}}{{Guilet} et~al.}{2013}]{2013MNRAS.430.1764G}
{Guilet} J.,  {Baruteau} C.,   {Papaloizou} J.~C.~B.,  2013, \mn@doi [\mnras]
  {10.1093/mnras/sts720}, \href
  {http://adsabs.harvard.edu/abs/2013MNRAS.430.1764G} {430, 1764}

\bibitem[\protect\citeauthoryear{{Ilgner} \& {Nelson}}{{Ilgner} \&
  {Nelson}}{2006}]{2006A&A...445..205I}
{Ilgner} M.,  {Nelson} R.~P.,  2006, \mn@doi [\aap]
  {10.1051/0004-6361:20053678}, \href
  {http://adsabs.harvard.edu/abs/2006A%26A...445..205I} {445, 205}

\bibitem[\protect\citeauthoryear{{Klahr} \& {Hubbard}}{{Klahr} \&
  {Hubbard}}{2014}]{2014ApJ...788...21K}
{Klahr} H.,  {Hubbard} A.,  2014, \mn@doi [\apj] {10.1088/0004-637X/788/1/21},
  \href {http://adsabs.harvard.edu/abs/2014ApJ...788...21K} {788, 21}

\bibitem[\protect\citeauthoryear{{Kley} \& {Nelson}}{{Kley} \&
  {Nelson}}{2012}]{2012ARA&A..50..211K}
{Kley} W.,  {Nelson} R.~P.,  2012, \mn@doi [\araa]
  {10.1146/annurev-astro-081811-125523}, \href
  {http://adsabs.harvard.edu/abs/2012ARA%26A..50..211K} {50, 211}

\bibitem[\protect\citeauthoryear{{Kunz}}{{Kunz}}{2008}]{2008MNRAS.385.1494K}
{Kunz} M.~W.,  2008, \mn@doi [\mnras] {10.1111/j.1365-2966.2008.12928.x}, \href
  {http://adsabs.harvard.edu/abs/2008MNRAS.385.1494K} {385, 1494}

\bibitem[\protect\citeauthoryear{{Kunz} \& {Lesur}}{{Kunz} \&
  {Lesur}}{2013}]{2013MNRAS.434.2295K}
{Kunz} M.~W.,  {Lesur} G.,  2013, \mn@doi [\mnras] {10.1093/mnras/stt1171},
  \href {http://adsabs.harvard.edu/abs/2013MNRAS.434.2295K} {434, 2295}

\bibitem[\protect\citeauthoryear{{Lesur}, {Kunz}  \& {Fromang}}{{Lesur}
  et~al.}{2014}]{2014A&A...566A..56L}
{Lesur} G.,  {Kunz} M.~W.,   {Fromang} S.,  2014, \mn@doi [\aap]
  {10.1051/0004-6361/201423660}, \href
  {http://adsabs.harvard.edu/abs/2014A%26A...566A..56L} {566, A56}

\bibitem[\protect\citeauthoryear{{Lin} \& {Youdin}}{{Lin} \&
  {Youdin}}{2015}]{2015ApJ...811...17L}
{Lin} M.-K.,  {Youdin} A.~N.,  2015, \mn@doi [\apj]
  {10.1088/0004-637X/811/1/17}, \href
  {http://adsabs.harvard.edu/abs/2015ApJ...811...17L} {811, 17}

\bibitem[\protect\citeauthoryear{{Lyra}}{{Lyra}}{2014}]{2014ApJ...789...77L}
{Lyra} W.,  2014, \mn@doi [\apj] {10.1088/0004-637X/789/1/77}, \href
  {http://adsabs.harvard.edu/abs/2014ApJ...789...77L} {789, 77}

\bibitem[\protect\citeauthoryear{{Masset}}{{Masset}}{2001}]{2001ApJ...558..453M}
{Masset} F.~S.,  2001, \mn@doi [\apj] {10.1086/322446}, \href
  {http://adsabs.harvard.edu/abs/2001ApJ...558..453M} {558, 453}

\bibitem[\protect\citeauthoryear{{Masset} \& {Ben{\'{\i}}tez-Llambay}}{{Masset}
  \& {Ben{\'{\i}}tez-Llambay}}{2016}]{2016ApJ...817...19M}
{Masset} F.~S.,  {Ben{\'{\i}}tez-Llambay} P.,  2016, \mn@doi [\apj]
  {10.3847/0004-637X/817/1/19}, \href
  {http://adsabs.harvard.edu/abs/2016ApJ...817...19M} {817, 19}

\bibitem[\protect\citeauthoryear{{Masset} \& {Papaloizou}}{{Masset} \&
  {Papaloizou}}{2003}]{2003ApJ...588..494M}
{Masset} F.~S.,  {Papaloizou} J.~C.~B.,  2003, \mn@doi [\apj] {10.1086/373892},
  \href {http://adsabs.harvard.edu/abs/2003ApJ...588..494M} {588, 494}

\bibitem[\protect\citeauthoryear{{Masset}, {D'Angelo}  \& {Kley}}{{Masset}
  et~al.}{2006}]{2006ApJ...652..730M}
{Masset} F.~S.,  {D'Angelo} G.,   {Kley} W.,  2006, \mn@doi [\apj]
  {10.1086/507515}, \href {http://adsabs.harvard.edu/abs/2006ApJ...652..730M}
  {652, 730}

\bibitem[\protect\citeauthoryear{{Meyer}, {Balsara}  \& {Aslam}}{{Meyer}
  et~al.}{2012}]{2012MNRAS.422.2102M}
{Meyer} C.~D.,  {Balsara} D.~S.,   {Aslam} T.~D.,  2012, \mn@doi [\mnras]
  {10.1111/j.1365-2966.2012.20744.x}, 422, 2102

\bibitem[\protect\citeauthoryear{{Mignone}, {Flock}, {Stute}, {Kolb}  \&
  {Muscianisi}}{{Mignone} et~al.}{2012}]{2012A&A...545A.152M}
{Mignone} A.,  {Flock} M.,  {Stute} M.,  {Kolb} S.~M.,   {Muscianisi} G.,
  2012, \mn@doi [\aap] {10.1051/0004-6361/201219557}, \href
  {http://adsabs.harvard.edu/abs/2012A%26A...545A.152M} {545, A152}

\bibitem[\protect\citeauthoryear{{Nelson}, {Gressel}  \& {Umurhan}}{{Nelson}
  et~al.}{2013}]{2013MNRAS.435.2610N}
{Nelson} R.~P.,  {Gressel} O.,   {Umurhan} O.~M.,  2013, \mn@doi [\mnras]
  {10.1093/mnras/stt1475}, \href
  {http://adsabs.harvard.edu/abs/2013MNRAS.435.2610N} {435, 2610}

\bibitem[\protect\citeauthoryear{{Paardekooper}}{{Paardekooper}}{2014}]{2014MNRAS.444.2031P}
{Paardekooper} S.-J.,  2014, \mn@doi [\mnras] {10.1093/mnras/stu1542}, \href
  {http://adsabs.harvard.edu/abs/2014MNRAS.444.2031P} {444, 2031}

\bibitem[\protect\citeauthoryear{{Paardekooper} \& {Mellema}}{{Paardekooper} \&
  {Mellema}}{2008}]{2008A&A...478..245P}
{Paardekooper} S.-J.,  {Mellema} G.,  2008, \mn@doi [\aap]
  {10.1051/0004-6361:20078592}, \href
  {http://adsabs.harvard.edu/abs/2008A%26A...478..245P} {478, 245}

\bibitem[\protect\citeauthoryear{{Paardekooper} \& {Papaloizou}}{{Paardekooper}
  \& {Papaloizou}}{2009}]{2009MNRAS.394.2283P}
{Paardekooper} S.-J.,  {Papaloizou} J.~C.~B.,  2009, \mn@doi [\mnras]
  {10.1111/j.1365-2966.2009.14511.x}, \href
  {http://adsabs.harvard.edu/abs/2009MNRAS.394.2283P} {394, 2283}

\bibitem[\protect\citeauthoryear{{Paardekooper}, {Baruteau}, {Crida}  \&
  {Kley}}{{Paardekooper} et~al.}{2010}]{2010MNRAS.401.1950P}
{Paardekooper} S.-J.,  {Baruteau} C.,  {Crida} A.,   {Kley} W.,  2010, \mn@doi
  [\mnras] {10.1111/j.1365-2966.2009.15782.x}, \href
  {http://adsabs.harvard.edu/abs/2010MNRAS.401.1950P} {401, 1950}

\bibitem[\protect\citeauthoryear{{Paardekooper}, {Baruteau}  \&
  {Kley}}{{Paardekooper} et~al.}{2011}]{2011MNRAS.410..293P}
{Paardekooper} S.-J.,  {Baruteau} C.,   {Kley} W.,  2011, \mn@doi [\mnras]
  {10.1111/j.1365-2966.2010.17442.x}, \href
  {http://adsabs.harvard.edu/abs/2011MNRAS.410..293P} {410, 293}

\bibitem[\protect\citeauthoryear{{Pandey} \& {Wardle}}{{Pandey} \&
  {Wardle}}{2008}]{2008MNRAS.385.2269P}
{Pandey} B.~P.,  {Wardle} M.,  2008, \mn@doi [\mnras]
  {10.1111/j.1365-2966.2008.12998.x}, \href
  {http://adsabs.harvard.edu/abs/2008MNRAS.385.2269P} {385, 2269}

\bibitem[\protect\citeauthoryear{{Pierens}}{{Pierens}}{2015}]{2015MNRAS.454.2003P}
{Pierens} A.,  2015, \mn@doi [\mnras] {10.1093/mnras/stv2024}, \href
  {http://adsabs.harvard.edu/abs/2015MNRAS.454.2003P} {454, 2003}

\bibitem[\protect\citeauthoryear{{Pierens} \& {Raymond}}{{Pierens} \&
  {Raymond}}{2016}]{2016MNRAS.462.4130P}
{Pierens} A.,  {Raymond} S.~N.,  2016, \mn@doi [\mnras]
  {10.1093/mnras/stw1904}, \href
  {http://adsabs.harvard.edu/abs/2016MNRAS.462.4130P} {462, 4130}

\bibitem[\protect\citeauthoryear{{Salmon}}{{Salmon}}{1998}]{salmon98}
{Salmon} R.,  1998, {Lectures on Geophysical Fluid Dynamics}.
Oxford

\bibitem[\protect\citeauthoryear{{Simon}, {Lesur}, {Kunz}  \&
  {Armitage}}{{Simon} et~al.}{2015}]{2015MNRAS.454.1117S}
{Simon} J.~B.,  {Lesur} G.,  {Kunz} M.~W.,   {Armitage} P.~J.,  2015, \mn@doi
  [\mnras] {10.1093/mnras/stv2070}, \href
  {http://adsabs.harvard.edu/abs/2015MNRAS.454.1117S} {454, 1117}

\bibitem[\protect\citeauthoryear{{Terquem}}{{Terquem}}{2003}]{2003MNRAS.341.1157T}
{Terquem} C.~E.~J.~M.~L.~J.,  2003, \mn@doi [\mnras]
  {10.1046/j.1365-8711.2003.06455.x}, \href
  {http://adsabs.harvard.edu/abs/2003MNRAS.341.1157T} {341, 1157}

\bibitem[\protect\citeauthoryear{{Urpin} \& {Brandenburg}}{{Urpin} \&
  {Brandenburg}}{1998}]{1998MNRAS.294..399U}
{Urpin} V.,  {Brandenburg} A.,  1998, \mn@doi [\mnras]
  {10.1046/j.1365-8711.1998.01118.x}, \href
  {http://adsabs.harvard.edu/abs/1998MNRAS.294..399U} {294, 399}

\bibitem[\protect\citeauthoryear{{Ward}}{{Ward}}{1991}]{1991LPI....22.1463W}
{Ward} W.~R.,  1991, in Lunar and Planetary Science Conference.

\bibitem[\protect\citeauthoryear{{Ward}}{{Ward}}{1992}]{1992LPI....23.1491W}
{Ward} W.~R.,  1992, in Lunar and Planetary Science Conference.

\bibitem[\protect\citeauthoryear{{Wardle} \& {Ng}}{{Wardle} \&
  {Ng}}{1999}]{1999MNRAS.303..239W}
{Wardle} M.,  {Ng} C.,  1999, \mn@doi [\mnras]
  {10.1046/j.1365-8711.1999.02211.x}, \href
  {http://adsabs.harvard.edu/abs/1999MNRAS.303..239W} {303, 239}

\bibitem[\protect\citeauthoryear{{Xu} \& {Bai}}{{Xu} \&
  {Bai}}{2016}]{2016ApJ...819...68X}
{Xu} R.,  {Bai} X.-N.,  2016, \mn@doi [\apj] {10.3847/0004-637X/819/1/68},
  \href {http://adsabs.harvard.edu/abs/2016ApJ...819...68X} {819, 68}

\bibitem[\protect\citeauthoryear{{Ziegler}}{{Ziegler}}{2004}]{2004JCoPh.196..393Z}
{Ziegler} U.,  2004, \mn@doi [\rm JCoPh] {10.1016/j.jcp.2003.11.003}, 196, 393

\bibitem[\protect\citeauthoryear{{de Val-Borro} et~al.,}{{de Val-Borro}
  et~al.}{2006}]{2006MNRAS.370..529D}
{de Val-Borro} M.,  et~al., 2006, \mn@doi [\mnras]
  {10.1111/j.1365-2966.2006.10488.x}, \href
  {http://adsabs.harvard.edu/abs/2006MNRAS.370..529D} {370, 529}

\makeatother
\end{thebibliography}



\appendix
\section{Vortensity Conservation and Slow Migration}
\label{sec:vortsource}

In this appendix we discuss the symmetry between the action of the body force in causing 
the gas to flow past the planet, and a radial migration of the planet past the gas.
This allows us to establish the conditions where we can appeal to vortensity conservation to derive an 
estimate of the corotation torque in analogy with the case of a moving planet.
First we transform the continuity and momentum equation into a time-dependent coordinate system, where 
the radial coordinate translates with time. 
This allows a vortensity-like quantity to be constructed, although one which has a source term 
related to the radial motion of the coordinate system.
Locally, this can be related to the Lorentz force generated by the laminar magnetic field.
We can then examine the conditions where the source term only leads to a small 
non-conservation of the vortensity-like quantity (or that this quantity and the actual vortensity are approximately equal).
This reduces to the condition that $\chi \gg 1$ or that the planet migration rate is slow.

We start with the continuity equation and momentum equation in two dimensions
\begin{align}
\frac{\partial}{\partial t} \Sigma + \nabla \cdot(\Sigma \V) & = 0\\
\frac{\partial}{\partial t} \V + (\V \cdot \nabla)\V + \frac{\nabla p}{\Sigma} &= -\nabla \Phi
\end{align}
where $\Phi$ is the gravitational potential and the gas is isothermal with $p= c_s^2 \Sigma$.
With a cylindrical coordinate system $(r,\phi)$ the velocity vector is $\V=(v_r, r\Omega)^T$.

To change to a radially moving coordinate system define a new radial coordinate
$r' = r/r_1(t)$ where the reference position $r_1 = r_0 + v_1 t$ with $v_1$ a constant velocity.
The cylindrical 
 gradient operator $\nabla$ becomes $\nabla'$ where $r$ is replaced by $r'$ and 
both components of the velocity $\U' = (u_r,r'\Omega)^T$ have units of angular velocity, with 
 \begin{align}
 u_r = \frac{v_r - r' v_1}{r_1}\, .
 \end{align}
 Defining $\Sigma' = r_1^2 \Sigma$, the continuity and momentum equations in the time-dependent coordinate system are:
 \begin{align}
 \frac{\partial'}{\partial t} \Sigma' + \nabla' \cdot(\Sigma' \U') & = 0 \\
 \frac{\partial'}{\partial t} \U' + (\U' \cdot \nabla')\U' + \frac{\nabla' p'}{\Sigma'} &= -\frac{\nabla' \Phi}{r_1^2} - 2\frac{v_1}{r_1} \U'
 \end{align}
These can now be used to form an equation
for the evolution of vortensity defined as $\xi' = (\nabla' \times \U')_z/\Sigma'$, which is
\begin{align}
\frac{\partial' }{\partial t} \xi' + \U ' \cdot \nabla' \xi' + 2 \frac{v_1 }{r_1} \xi' = 0 \, .
\label{eqVort}
\end{align}
Vortensity is therefore not conserved along streamlines in this coordinate frame because of the source term. The importance of the source term can be established by noting that it can be removed 
by transforming into the quantity
$\xi'' = \xi'(1+v_1 t/r_0)^2$
which has the evolution equation
\begin{align}
\frac{\partial' }{\partial t} \xi'' + \U ' \cdot \nabla' \xi'' = 0\, .
\end{align}
The vortensity-like quantity $\xi''$ is thus conserved.
Hence we are interested in a condition for $\xi' \approx \xi''$ on the relevant scales for the corotation torque, since in this case we can take the source term in (\ref{eqVort}) to be unimportant. 
From expanding $\xi''$ in powers of $v_1 t/r_0$ we can see that radially, on the scale of 
the corotation region $\sim 2 x_s$ the fractional change in $\xi''$ is $\sim 2 x_s/r_1$.
Thus we want $ 2 x_s/r_1 \ll 2 v_1 t/r_0$. If we take the time-scale for the motion on the right hand side of this expression 
to be the libration time-scale $\tau_{\rm lib} \sim r_0/(\Omega x_s)$ we find that it reduces to the $\chi\gg 1$ condition,
 or equivalently the slow migration condition as appears in \citet{2003ApJ...588..494M} and \citet{2014MNRAS.444.2031P}. Therefore, when $\chi\gg 1$, we can take vortensity to be conserved both in the case of a migrating planet in a non-accreting disc and a static planet in a disc with an accretion flow past the planet.

\section{Resolution Study}
\label{sec:resolutionstudy}

\begin{figure}
\includegraphics[width=\columnwidth]{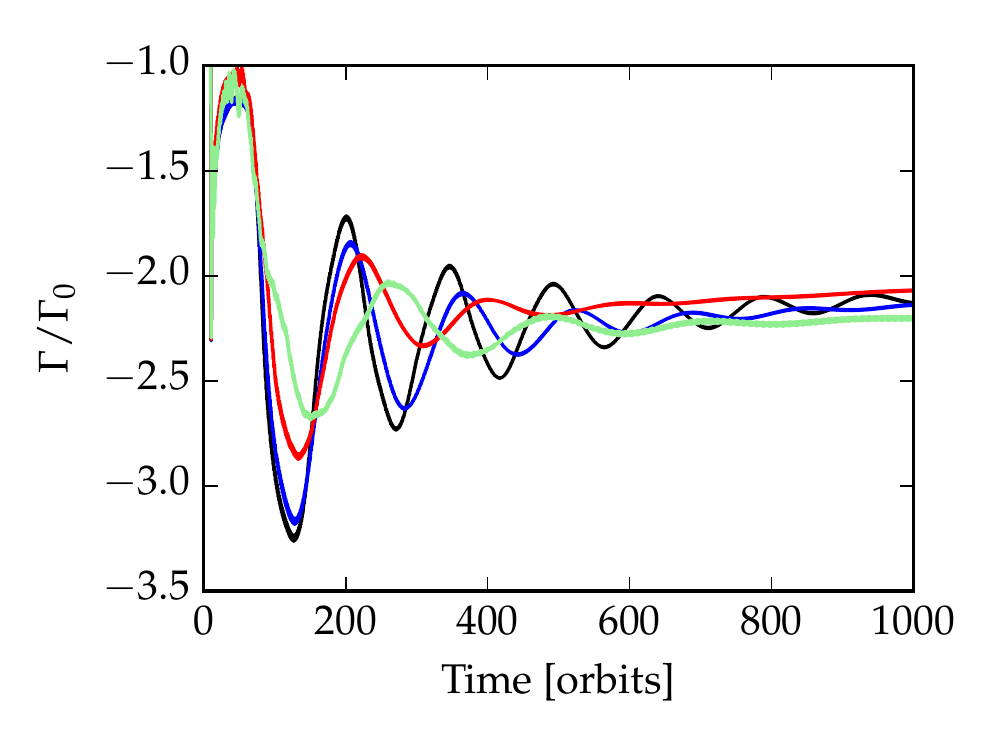}
\caption{Torque histories for the $\chi = 10\sqrt{2}$ at varying resolutions.
Grid resolutions are
Light green: $N_r, N_\phi = [256,768]$ Red: $N_r, N_\phi = [512,1536]$ Blue: $N_r, N_\phi = [1024,3072]$ Black: $N_r, N_\phi = [2048,6144]$
}
\label{fig:resstudy}
\end{figure}

Fig.~\ref{fig:resstudy} shows the resolution dependence of the Lindblad torque, and damping of the libration oscillations, and 
the gradual increase in the slope of the growing corotation torque due to the vortensity growth of the libration region.
This gradually steepening growth of the corotation toque is due to the lesser degree of numerical diffusion smearing the vortensity jump at the 
edge of the corotation region.


\bsp	
\label{lastpage}
\end{document}